\documentclass[aps,prb,twocolumn]{revtex4}
\usepackage{boxedminipage}
\usepackage{lscape}
\usepackage{float}
\usepackage[ansinew]{inputenc}
\usepackage{epsfig}
\usepackage{subfigure}
\usepackage{graphicx}
\usepackage{amsmath}
\usepackage{amssymb}
\usepackage{amsthm,url}
\usepackage[french]{babel}  

\renewcommand{\d}[2]{\frac{#1}{#2}}
\newcommand{\pd}{\partial}

\newcommand{\barray}{\begin{eqnarray}}
\newcommand{\earray}{\end{eqnarray}}

\newcommand{\beq}{\begin{equation}}

\newcommand{\eeq}{\end{equation}}

\begin{document}

\title{Benchmark of a modified Iterated Perturbation Theory approach on the FCC lattice at strong coupling}
\author{Louis-Fran\c{c}ois Arsenault$^{1}$, Patrick S\'{e}mon$^{1}$, A.-M. S. Tremblay$^{1,2}$}
\affiliation{$^1$ D\'{e}partement de Physique and RQMP, Universit\'{e} de Sherbrooke, Sherbrooke,
QC, Canada\\
$^{2}$Canadian Institute for Advanced Research, Toronto, Ontario, Canada.}
\date{\today}

\begin{abstract}
The Dynamical Mean-Field theory (DMFT) approach to the Hubbard model requires a method to solve the problem of a quantum impurity in a bath of non-interacting electrons. Iterated Perturbation Theory (IPT) has proven its effectiveness as a solver in many cases of interest. Based on general principles and on comparisons with an essentially exact Continuous-Time Quantum Monte Carlo (CTQMC) solver, here we show that the standard implementation of IPT fails away from half-filling when the interaction strength is much larger than the bandwidth. We propose a slight modification to the IPT algorithm that replaces one of the equations by the requirement that double occupancy calculated with IPT gives the correct value. We call this method IPT-$D$. We recover the Fermi liquid ground state away from half-filling. The Fermi liquid parameters, density of states, chemical potential, energy and specific heat on the FCC lattice are calculated with both IPT-$D$ and CTQMC as benchmark examples. We also calculated the resistivity and the optical conductivity within IPT-$D$. Particle-hole asymmetry persists even at coupling twice the bandwidth. Several algorithms that speed up the calculations are described in appendices.
\end{abstract}

\maketitle

\hyphenation{Brillouin}

\section{Introduction}
Within the last fifteen years or so, the Dynamical mean field theory approach (DMFT)\cite{Georges:1992,Jarrell:1992,Georges:1996} and its cluster generalizations\cite{Hettler:1998,Kotliar:2001,Maier:2005,KotliarRMP:2006} have become some of the most powerful techniques to study strongly correlated electrons. In these approaches, a single-site hybridized to a bath or a cluster hybridized to a bath must be solved. The bath of non-interacting electrons is determined self-consistently. At the heart of the DMFT approach then, one finds so-called impurity solvers. There are now very powerful impurity solvers, for example Continuous Time Quantum Monte Carlo (CTQMC) methods\cite{Gull:2011}. These methods are exact within statistical errors and, for the one-band Hubbard model, certain versions\cite{WernerMillis:2006} at the single site level do not suffer from sign problem. Yet, these approaches require sizeable computational resources and, in addition, real frequency information must be obtained through analytical continuation of data with statistical uncertainties, an ill-posed problem\cite{Jarrell:1996}. It is thus still of great interest to work with approximate solvers that are reliable and do not suffer from statistical uncertainties. This facilitates the calculation of real-frequency quantities with Pad\'e approximants\cite{Vidberg:1977} or directly in real-frequency and also enables one to quickly explore phase diagrams and pinpoint interesting regions of parameter space where state of the art solvers would be useful. Among possible approximate solvers, one finds exact diagonalization, slave bosons, Non-Crossing Approximation (NCA), Numerical Renormalization Group (NRG) and others\cite{Georges:1996}. They all have advantages and disadvantages. For example, exact diagonalization can consider only a limited number of bath sites, NCA is limited to high temperatures and NRG to low energies.

Here we consider Iterated Perturbation Theory (IPT) \cite{Kajueter:1996}, an interpolation approach that generalizes the original\cite{Rozenberg:1994,Rozenberg:1995} IPT applicable only at half filling. This method has been, and still is, wildly used\cite{KotliarRMP:2006,Held:2007}. The interpolation is constructed so that the self-energy recovers both the exact result in the atomic limit and the high-frequency limit of the Hubbard model. There is one parameter however that cannot be determined from these constraints. There have been several proposals to fix this parameter. At $T=0$ one can impose that Luttinger's theorem be satisfied (IPT-$L$) as was done in [\onlinecite{Kajueter:1996}] but when this condition is applied at finite temperature, the results are not satisfactory\cite{PhDKajueter}. Another very popular approach for non-zero temperature fixes the occupation $n_0$ of the non-interacting part of the Anderson impurity problem used in the perturbative calculation to be equal to the lattice occupation $n=n_0$ \cite{Potthoff:1997,Martin-Rodero:1986} (IPT-$n_0$). This condition is arbitrary since there is no general principle relating these two numbers, but it turns out to be quite satisfactory in the case of correlated metal i.e. $U < U_{Mott}$\cite{Potthoff:1997,Meyer:1999,Merino:2000}. $U_{Mott}$ is the coupling for which the metal to insulator Mott transition occurs at half-filling.

Despite this success, it is known\cite{Potthoff:1997} that when $U$ is larger than the critical value for the Mott transition at $n=1$ ($U>U_{Mott}$), then IPT breaks down for $n>1$ at low $T$ when the condition $n=n_0$ is applied. This happens even if, in principle, IPT is constructed to respect the atomic limit $U \gg t$. It has been proposed\cite{Potthoff:1997} that preserving the third moment of the spectral weight improves the results. Here we show that IPT-$n_0$ is unsatisfactory for $U\gg U_{Mott}$ close to half-filling for both $n>1$ and $n<1$. We propose a way to circumvent this problem by using the fact that when $U$ is large enough, the double occupancy becomes almost temperature independent in the paramagnetic state with a value that is, to a high degree of accuracy, a simple function of the density. This provides us with a condition different from $n=n_0$ that allows one to close the IPT equations even for large coupling. This approach, IPT-$D$, is applicable at all temperatures contrary to the approach that enforces Luttinger's theorem. It can in principle be improved further by enforcing the third-moment sum rule\cite{Potthoff:1997}.

In Sec.~\ref{DMFT} we summarize the DMFT approach, the solvers that we use and the manner in which Fermi liquid parameters are extracted. Sec.~\ref{breakdown} demonstrates the failure of IPT at large coupling. In this section and throughout the text, numerical examples are obtained with the 3-dimensional FCC lattice. Amongst lattice presenting electronic frustration, the FCC lattice is important because of its prevalence in nature. Our main contribution appears in Sec.~\ref{Sec:Double} where we show that double occupancy can be accurately determined from simple arguments at very strong coupling and then used to fix the remaining parameter in IPT. We call this approach IPT-$D$. Fermi liquid parameters, density of states, chemical potential, energy, and specific heat on the FCC lattice are calculated with both IPT-$D$ and CTQMC as benchmark examples. Resistivity and optical conductivity obtained with IPT-$D$ are physically reasonable. Appendix A contains details on the three dimensional adaptive integrator we developed for both IPT and CTQMC calculation.  Appendix B contains details of the implementation of IPT-$D$. Appendix C explains how to calculate the different non-interacting functions and Appendix D gives details on the calculation of the optical conductivity.

\section{Model, DMFT and impurity solver}\label{DMFT}

\subsection{Model and DMFT}
We study the one-band Hubbard model,
\begin{equation}
H=-\sum_{i,j,\sigma }t_{i,j}d_{i,\sigma }^{\dagger }d_{j,\sigma
}+U\sum_{i}n_{i\uparrow }n_{i\downarrow }
\end{equation}%
where $t_{i,j}$ is the hopping matrix between sites $i$ and $j$, $U$ the on-site Coulomb repulsion, $d_{i,\sigma }^{(\dagger )}$ the creation (annihilation) operator for an electron of spin $\sigma$ on site $i$ and $n_{i\sigma }=d_{i,\sigma }^{\dagger }d_{i,\sigma }$ is the number operator.

The dynamical mean-field theory (DMFT) provides a solution of the Hubbard model that describes the Mott transition in three dimensions and has predictive power for real materials\cite{KotliarRMP:2006,Georges:1996}. Drawing from ideas on the solution of the Hubbard model in infinite dimension\cite{Metzner:1989}, the self-energy in this approach depends only on frequency. One first solves the problem of a single site with the Hubbard $U$, hybridized with an infinite bath of non-interacting electrons, the so-called Anderson model. One extracts the frequency-dependent self-energy of the Anderson model, which is then taken as the self-energy in the lattice Green's function. The bath is determined self-consistently by requiring that projection of the lattice Green's function on a single site is identical to the single-site Green's function of the Anderson model. The Anderson impurity problem can be solved numerically with a very high precision. DMFT has been justified with a variety of approaches\cite{Georges:1996} including a variational one\cite{Potthoff:2003b}. The single-site DMFT is exact in infinite dimension\cite{Georges:1996}. Benchmarks against the Bethe ansatz solution in one-dimension shows that DMFT can be an accurate solution of the Hubbard model also in lower dimensions \cite{Bolech:2003,Capone:2004}.

\begin{figure}[tbp]
\begin{center}
\includegraphics[width=0.9\linewidth]{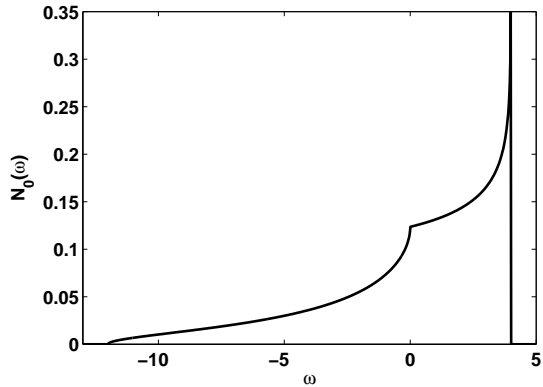}
\end{center}
\caption{Non-interacting density of states for the FCC lattice with nearest-neighbor hopping only. The large particle-hole asymmetry caused by frustration is apparent.}
\label{fig:DOS_U0}
\end{figure}

Mathematically, the partition function for the Anderson impurity problem is given by the imaginary-time Grassmann path integral
\begin{equation}
Z=\int \mathcal{D}[\psi ^{\dag },\psi ]\,\mathrm{e}^{-S_{0}-\int_{0}^{\beta
}d\tau \int_{0}^{\beta }d\tau ^{\prime \dag }(\tau )\Delta (\tau ,\tau
^{\prime })\psi (\tau ^{\prime })},  \label{eq:Z}
\end{equation}%
with $\hbar =1,\beta ^{-1}=k_{B}T,$ $\Delta $ the bath hybridization and $%
S_{0}$ the action of the impurity, which consists of a single site with
repulsion $U$. The self-consistency condition in Matsubara frequency reads
\begin{equation}
\begin{split}
\Delta (i\omega _{n})=& \,i\omega _{n}+\mu -\Sigma (i\omega _{n}) \\
& \,-\left[ \sum_{\mathbf{k}}\frac{1}{i\omega _{n}+\mu -\varepsilon _{%
\mathbf{k}}-\Sigma (i\omega _{n})}\right] ^{-1}.
\end{split}
\label{eq:SCC}
\end{equation}%
with $\Sigma $ the self-energy. On the FCC lattice the single-particle
dispersion is given, with lattice spacing $a=1$, by $\varepsilon _{\mathbf{k}}=-4t[\cos (k_{x})\cos(k_{y})+\cos (k_{x})\cos (k_{z})+\cos (k_{y})\cos (k_{z})]$. In single-site DMFT, the self-energy is local ($\Sigma (\omega)$) which enables one to transform the integrals over the Brillouin zone entering the self-consistency relation into integrals over the non-interacting density of states $N_0(\omega)$. However, in the case of a 3d FCC lattice, there is no analytic form for $N_0(\omega)$ and its accurate numerical calculation using a Monte-Carlo binning procedure is cumbersome and only produces a fixed finite numbers of points. On the other hand, if we used a Lorentzian as an approximation for the delta function, the band-edges and the Van-Hove singularities would suffer from accuracy problems that could be transferred to the DMFT calculation. We thus performed the calculation with the full $k$ space integration. We devised an adaptive 3d fifth order Gaussian quadrature for a cube. This integration method is explained in Appendix~\ref{appen_integrate}. For comparisons in calculation of transport properties, we have nevertheless calculated $N_0(\omega)$ using Monte Carlo integration as explained in Appendix~\ref{MC_function}. The resulting non-interacting density of state for the FCC lattice with nearest-neighbor hopping only is shown in Fig.~\ref{fig:DOS_U0}.

We have used two "impurity solvers" for the auxiliary Anderson model. They are described in the following subsections.

\subsection{CTQMC}
The first method is the numerically exact continuous time quantum Monte Carlo method (CTQMC) \cite{Werner:2006}, a finite temperature approach that relies on the Monte Carlo summation of all diagrams obtained from the expansion of the partition
function in powers of the hybridization $\Delta$. This method does not have a sign problem, and does not have errors associated with time discretization or bath parametrization. It is therefore exact within statistical errors but computationally expensive. We refer to the literature for an explanation of the approach\cite{WernerMillis:2006,Gull:2011}.

\subsection{IPT}
We describe the second approach, Iterated Perturbation Theory (IPT), in more details since it is the focus of this paper. IPT is an approximation method that relies on an interpolation from $2^{nd}$ order perturbation
theory for the Anderson impurity problem \cite{Kajueter:1996}. The interpolation preserves the correct high-frequency limit for the self-energy
and is exact in both the non-interacting and the atomic limits. We only consider paramagnetic solutions.

The self-energy in this approach is parametrized by
\begin{equation}\label{IPT_self}
    \Sigma (i\omega_n) = U\d{n}{2} + \d{A\Sigma^{(2)}(i\omega_n)}{1-B\Sigma^{(2)}(i\omega_n)},
\end{equation}
where
\begin{equation}\label{self_2order_AIM}
    \Sigma^{(2)}(i\omega_n) = U^2\int_0^{\beta}d\tau\text{e}^{i\omega_n\tau}G_0^{\sigma}(\tau)G_0^{-\sigma}(-\tau)G_0^{-\sigma}(\tau),
\end{equation}
with
\begin{equation}\label{G0}
    G_0(i\omega_n) = \d{1}{i\omega_n + \mu_0 - \Delta (i\omega_n)}
\end{equation}
and $\Delta$ the hybridization function. The constants $A$ and $B$
\begin{equation}\label{AB}
\begin{split}
    A &= \d{n(2-n)}{n_0(2-n_0)}\\
    B &= \d{(1-\d{n}{2})U + \mu_0 - \mu}{\d{n_0}{2}(1-\d{n_0}{2})U^2},
\end{split}
\end{equation}
where $n_0 = 2G_0(\tau = 0^-)$ and $n = 2G(\tau = 0^-)$, are chosen such that one recovers the exact solution in the atomic limit as well as the exact result for arbitrary $U$ in the high-frequency limit. The Green's function used to obtain the density $n$ is
\begin{equation}\label{G_loc_def}
    G(i\omega_n) = \sum_k\d{1}{i\omega_n - (\varepsilon_k-\mu)-\Sigma (i\omega_n)}.
\end{equation}
In Eq.~\eqref{AB}, $\mu$ is the chemical potential of the lattice that is determined by fixing the value of $n$ while $\mu_0$ is the chemical potential determined by the fictitious density $n_0$.

We need an additional equation to fix $\mu_0$. This problem has been studied carefully in Refs.[\onlinecite{Potthoff:1997,Meyer:1999}]. Setting $\mu=\mu_0$ is not a good option. Indeed, as mentioned in the introduction, fixing Luttinger's volume works only at very low temperature\cite{PhDKajueter}. A widely used approach~\cite{Martin-Rodero:1986} consists in fixing $n=n_0$. We call this approach IPT-$n_0$. One can also modify the formula for the interpolated self-energy by requiring that the third moment, appearing in the high-frequency expansion of the Green's function, be satisfied exactly. In this case, the deficiencies of IPT-$n_0$ at strong coupling are not as severe. We will see below that requiring that double-occupancy takes its exact value is an easier solution that does not modify the simplicity of the original scheme and gives accurate results.

IPT can be implemented efficiently, as described in Appendix.~\ref{appen_IPT}, so that the solution can be obtained in a very short time.

\section{Breakdown of IPT}\label{breakdown}

In this section, we first define the physical parameters that will be used to demonstrate the breakdown of IPT-$n_0$. Then we take advantage of the existence of the exact CTQMC impurity solver to characterize the solution of the DMFT equation. The last subsection demonstrates that for $U$ much larger than the bandwidth, IPT-$n_0$ fails to reproduce even qualitatively the exact solution.

\subsection{Extracting the Fermi liquid parameters}
At low $T$ and finite doping, it is known that DMFT predicts a Fermi liquid regime no matter how strong the interaction\cite{Georges:1996}. In other words, a quasiparticle peak always appears at $\omega = 0$ at low $T$ except at half-filling when $U > U_{Mott}$. We characterize the Fermi liquid with three parameters namely the effective chemical potential,
\begin{equation}
\tilde{\mu} = \mu - \Sigma'(0),
\end{equation}
the quasiparticle weight,
\begin{equation}
Z = \left(1-\d{\pd\Sigma'(\omega)}{\pd\omega}\Big|_{\omega\rightarrow 0}\right)^{-1}
\end{equation}
and the scattering rate $\Sigma''(0)$, where real and imaginary parts are defined by  $\Sigma = \Sigma'+i\Sigma''$.
In the DMFT treatment of the Hubbard model, Luttinger's theorem is satisfied at $T=0$ when $\tilde{\mu}$ takes the value of the non-interacting chemical potential that gives the same density.

All of the above parameters are calculated with the self-energy on the real frequency axis and thus, in principle, one needs to perform an analytical continuation from the data in Matsubara frequencies and then extrapolate to zero temperature. In practice, we calculate the values of the self-energy for a few very low temperatures and use them to extrapolate to zero frequency and zero temperature. For the retarded $\Sigma'(0)$ and $\Sigma''(0)$ we take $\Sigma (\omega_{n=0})$ at the smallest positive $\omega_{n=0}$ for three low temperatures and extrapolate to $T=0$ using the fact that $\Sigma (\omega_n \rightarrow 0^+) = \Sigma^R (\omega \rightarrow 0)$. For $Z$, the spectral definitions of the self-energies
\begin{equation}\label{spectral_self}
\begin{split}
    \Sigma'(\omega) &= P\int\d{d\omega'}{\pi}\d{\Sigma''(\omega')}{\omega'-\omega}\\
    \Sigma(i\omega_n) &= \int\d{d\omega'}{\pi}\d{\Sigma''(\omega')}{\omega'-i\omega_n},
\end{split}
\end{equation}
allow one to prove
\begin{equation}\label{Z_matsu}
    \d{\textrm{Im}[\Sigma(i\omega_n)]}{\omega_n}\Big|_{\omega_n\rightarrow 0} = \int\d{d\omega'}{\pi}\d{\Sigma(\omega')}{\omega'^2} = \d{\pd\Sigma'(\omega)}{\pd\omega}\Big|_{\omega = 0}
\end{equation}
which, for a linear dependence of $\textrm{Im}[\Sigma(i\omega_n)]$ on $\omega_n$, also follows from the Cauchy-Riemann relation for holomorphic functions of a complex variable. This last equation with three low temperatures allows us to calculate $\textrm{Im}[\Sigma(i\omega_{n=0})]/\omega_{n=0}$, extrapolate to $T=0$ and obtain $Z$.

\subsection{Expected behavior, as obtained from CTQMC}

\begin{figure}[tbp]
\begin{center}
\mbox{\includegraphics[width=0.9\linewidth]{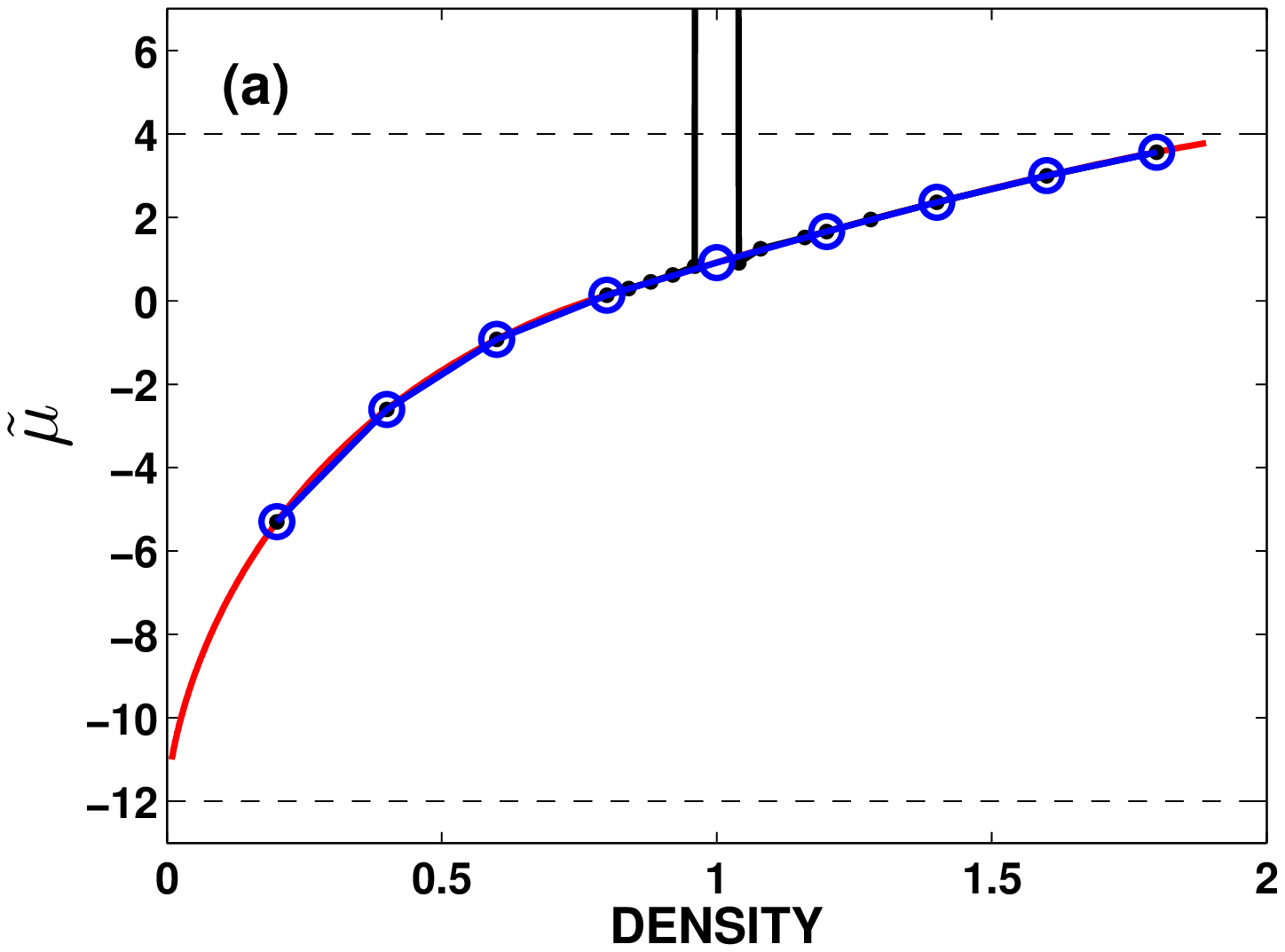}}
\mbox{\includegraphics[width=0.9\linewidth]{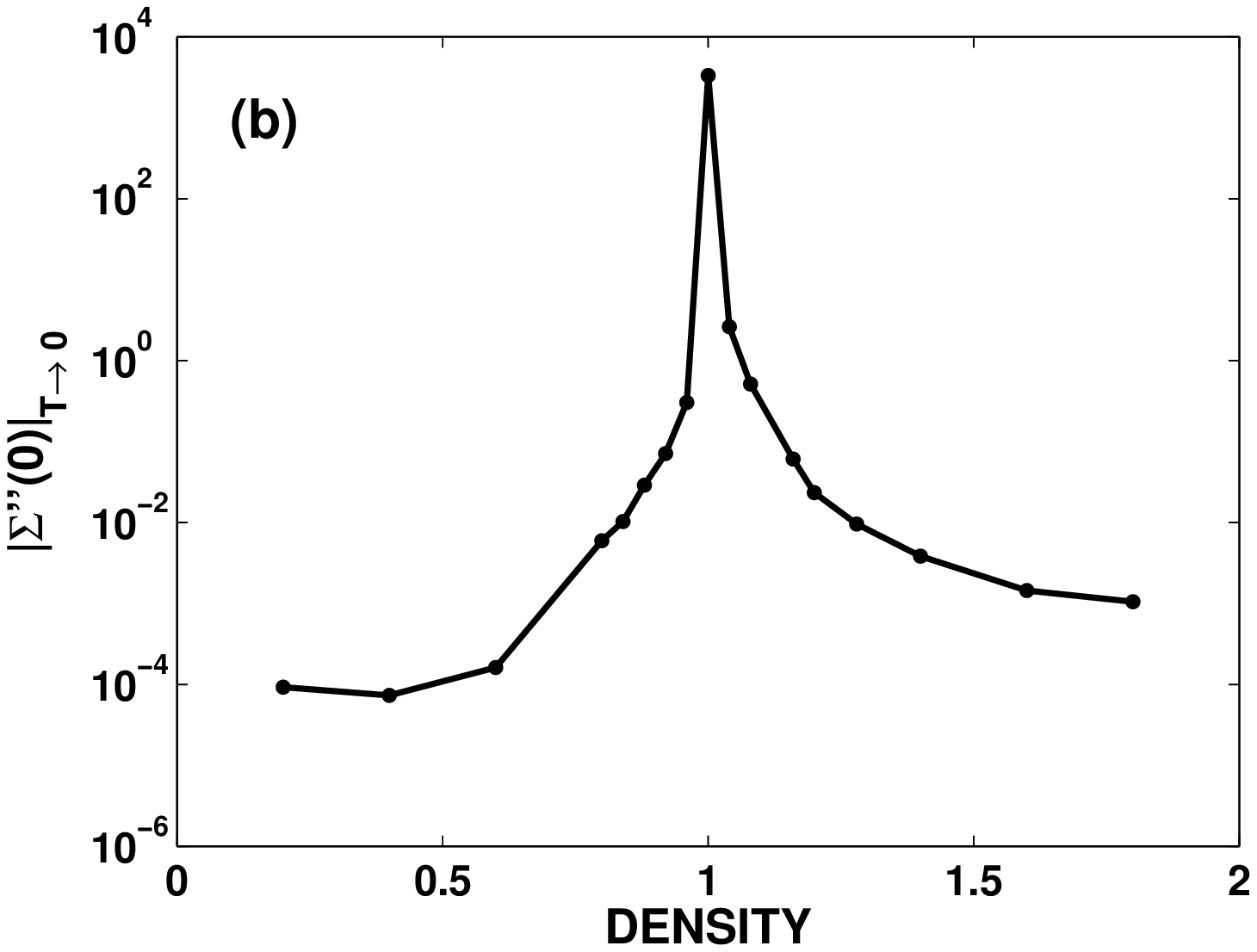}}
\mbox{\includegraphics[width=0.9\linewidth]{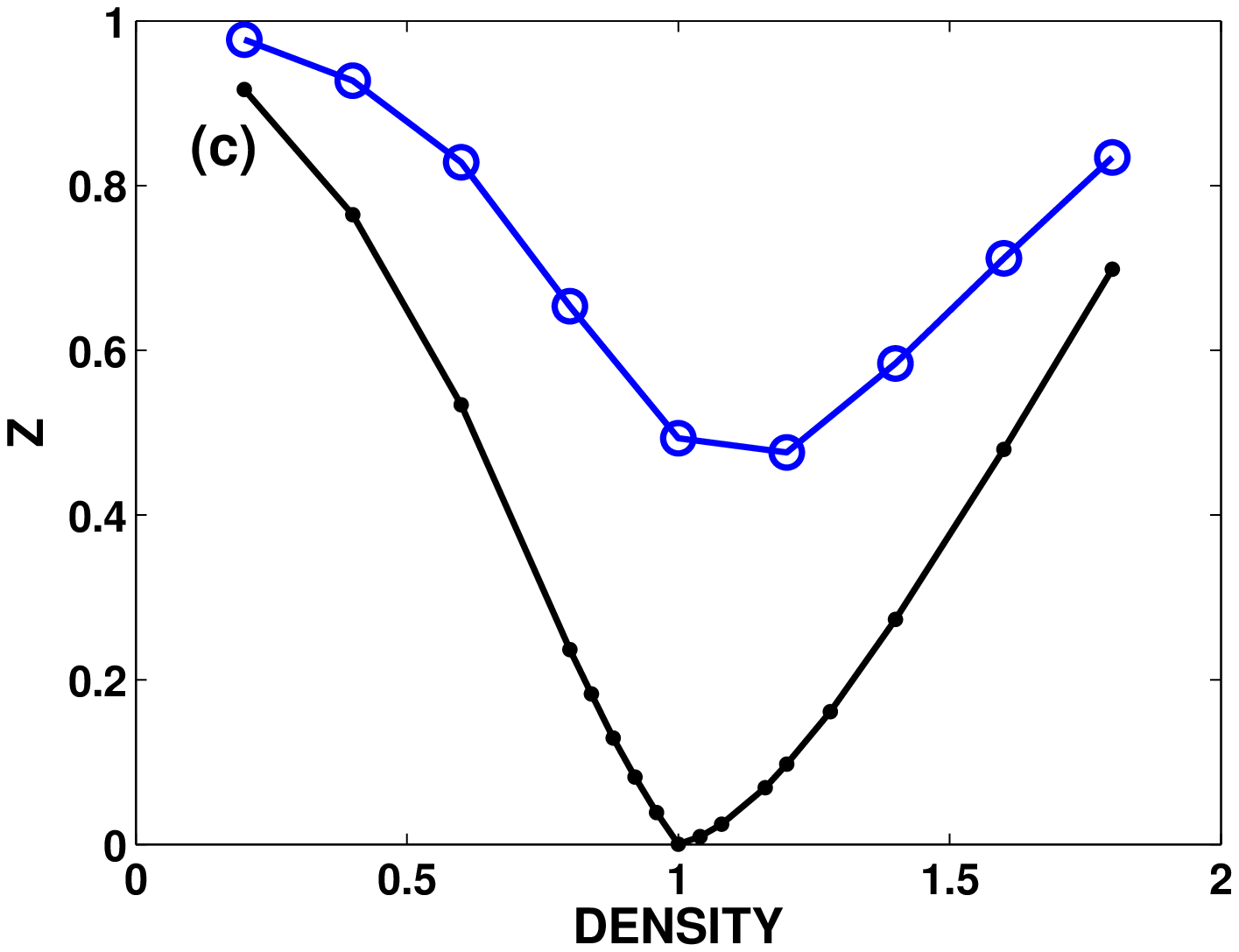}}
\end{center}
\caption{(Color online)Results obtained with CTQMC as impurity solver are plotted as a function of density and shown in blue with circles and line for $U=8$ and in black with dots and line for $U=32t$. In all numerical results, energy units are such that $t=1$. Boltzmann's constant and the lattice spacing are also taken as unity. We obtain the zero-frequency limit from a poor man's approach: we take $\beta t = 25, 50$ and $75$ and use the value of the function at the lowest Matsubara frequency in the three cases to perform the extrapolation. (a) Check for Luttinger's theorem: The effective chemical potential $\tilde{\mu}= \mu - \Sigma'(0)$ is equal to the non-interacting chemical potential shown in red except at half-filling where there is a Mott gap for $U=32t$. (b) At $U=32t$ the imaginary part of the self-energy at zero frequency $\Sigma''(0)$ should be zero away from half-filling and infinite at half-filling. (c) The single-particle spectral weight $Z$ vanishes only at $n=1$, $U=32t$ where there is a Mott gap.}
\label{fig:Fermi_CTQMC}
\end{figure}

Consider the one-band Hubbard model on the FCC lattice, where the single particle dispersion is given by $\varepsilon _{\mathbf{k}}=-4t[\cos (k_{x})\cos(k_{y})+\cos (k_{x})\cos (k_{z})+\cos (k_{y})\cos (k_{z})]$. We present the DMFT results obtained with the CTQMC impurity solver for $U=8t$ and $U=32t$, below and above  $U_{Mott}$ for the Mott transition at half-filling. The bandwidth is $16t$ for the 3d FCC lattice.

Fig.\ref{fig:Fermi_CTQMC} displays the Fermi liquid parameters. In Fig.\ref{fig:Fermi_CTQMC}-(a) the red solid line shows the non-interacting chemical potential as a function of density. The effective chemical potential $\tilde{\mu} = \mu - \Sigma'(0)$ is shown with blue circles for $U=8t$ and black points for $U=32t$. The dashed lines indicates the position of the band edges for the 3d FCC lattice. As expected, except at half-filling for $U=32t$, Luttinger's theorem is satisfied. In Fig.\ref{fig:Fermi_CTQMC}-(b), the extrapolated scattering rate $\Sigma''(\omega)$ is negligibly small, except for $U=32t$ at $n=1$. For $U=8t$, (not shown) it is of the order $1\text{x}10^{-4}$ and has essentially no density dependance. The value of $Z$, shown in In Fig.\ref{fig:Fermi_CTQMC}-(c), behaves as expected: For $U>U_{Mott}$, $Z$ vanishes when the occupation approaches half-filling while it is close to the non-interacting value $Z=1$ when the lattice is almost empty or full. We can also see that even for coupling as large as $U=32t$, the absence of particle-hole symmetry in the dispersion relation still leads to a value of $Z$ that is not symmetric with respect to half-filling. Clearly, electronic frustration plays an important role in the doped Mott insulator.

\subsection{Breakdown of IPT-$n_0$ }
The IPT equations Eqs.~\eqref{IPT_self}-\eqref{G_loc_def} do not determine the value of $n_0$. As mentioned previously, for $T=0$ the requirement that Luttinger's theorem be satisfied (IPT-$L$) provides an additional independent equation, except at half-filling for $U>U_{Mott}$. However, Luttinger's theorem is in general not satisfied at finite temperature and the method becomes inaccurate. The condition $n=n_0$ has thus been proposed\cite{Potthoff:1997,Martin-Rodero:1986}  (IPT-$n_0$). It gives satisfactory results for correlated metals (not for the insulator at half-filling).

\begin{figure}[tbp]
\begin{center}
\mbox{\includegraphics[width=0.9\linewidth]{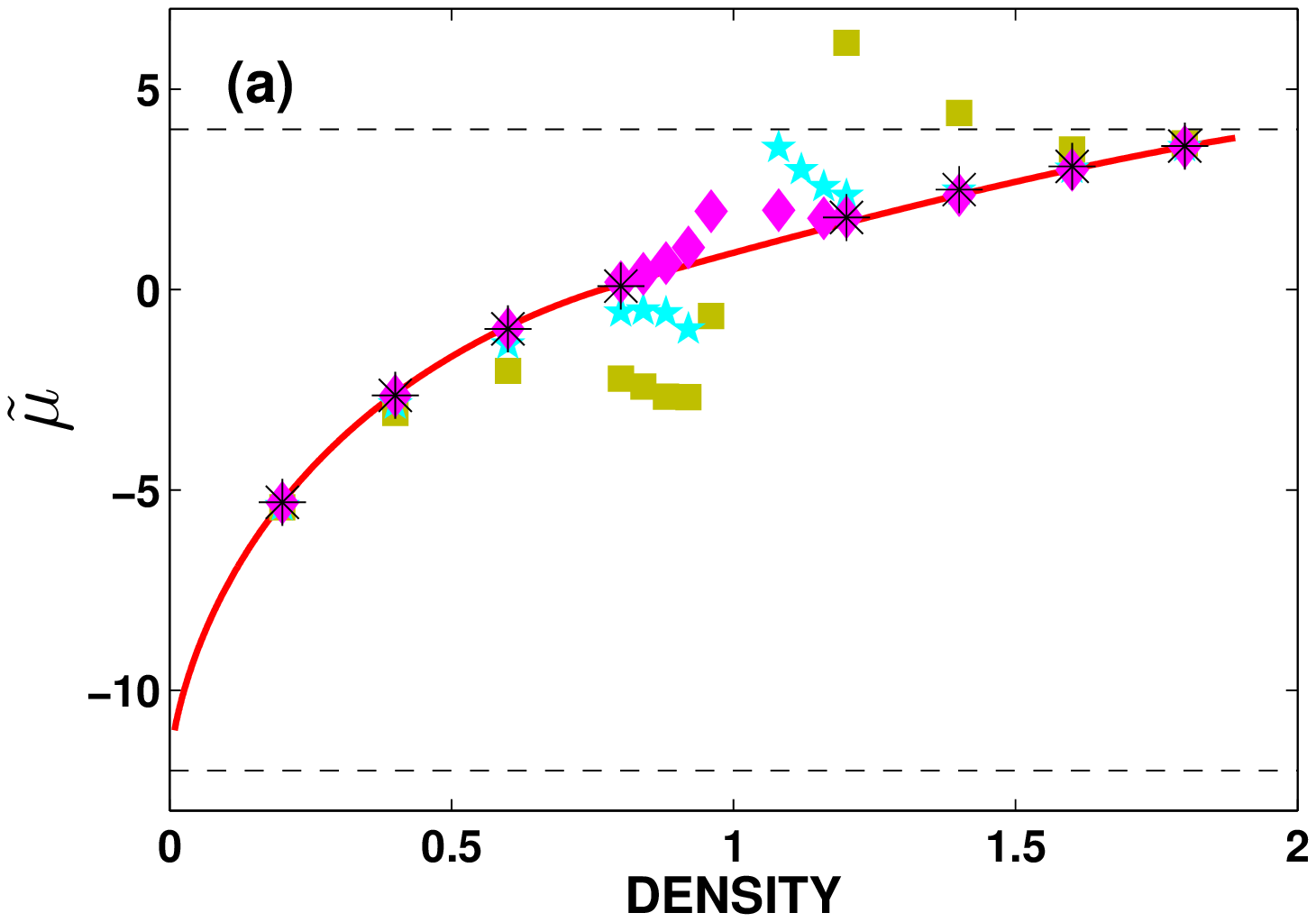}}
\mbox{\includegraphics[width=0.9\linewidth]{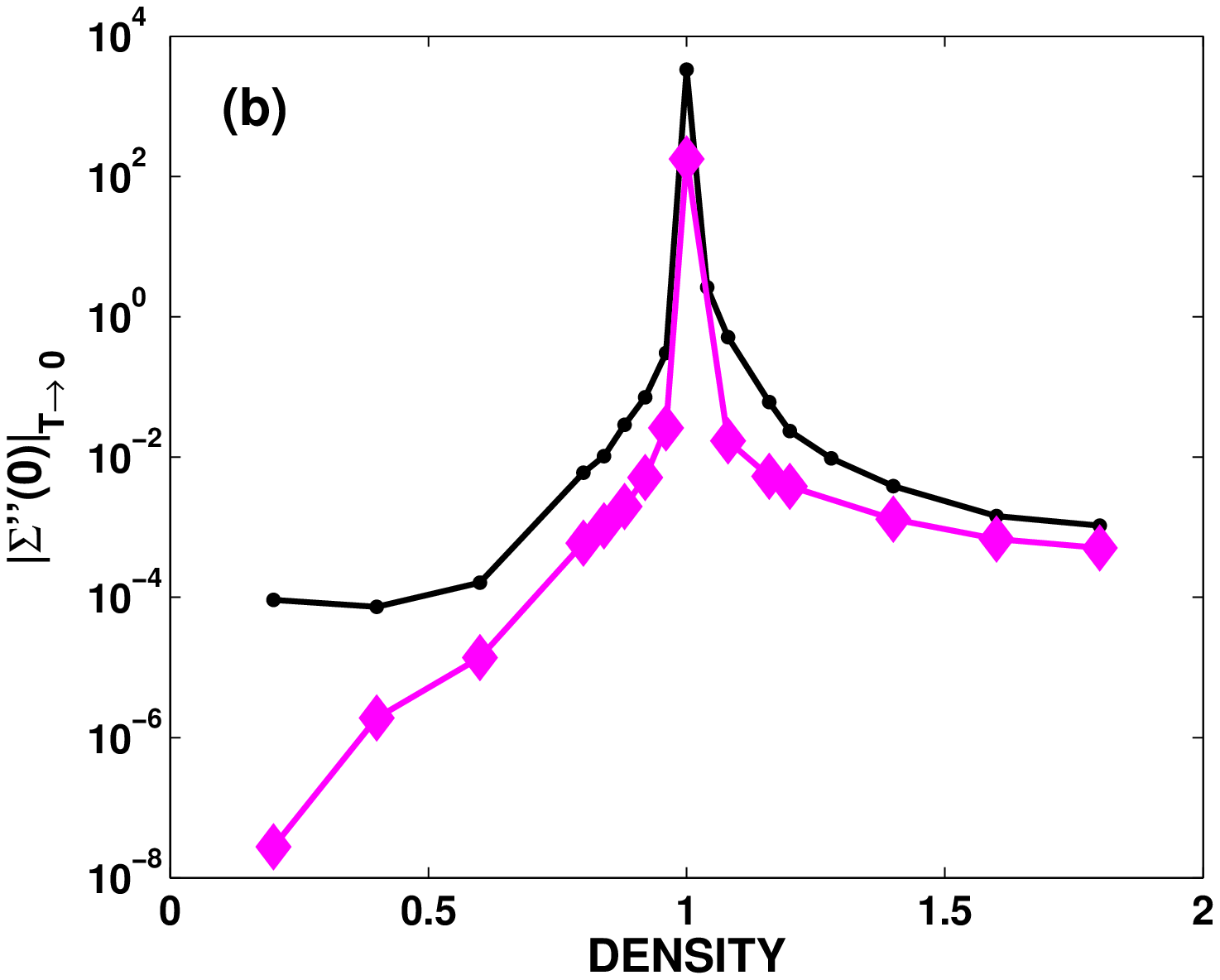}}
\mbox{\includegraphics[width=0.9\linewidth]{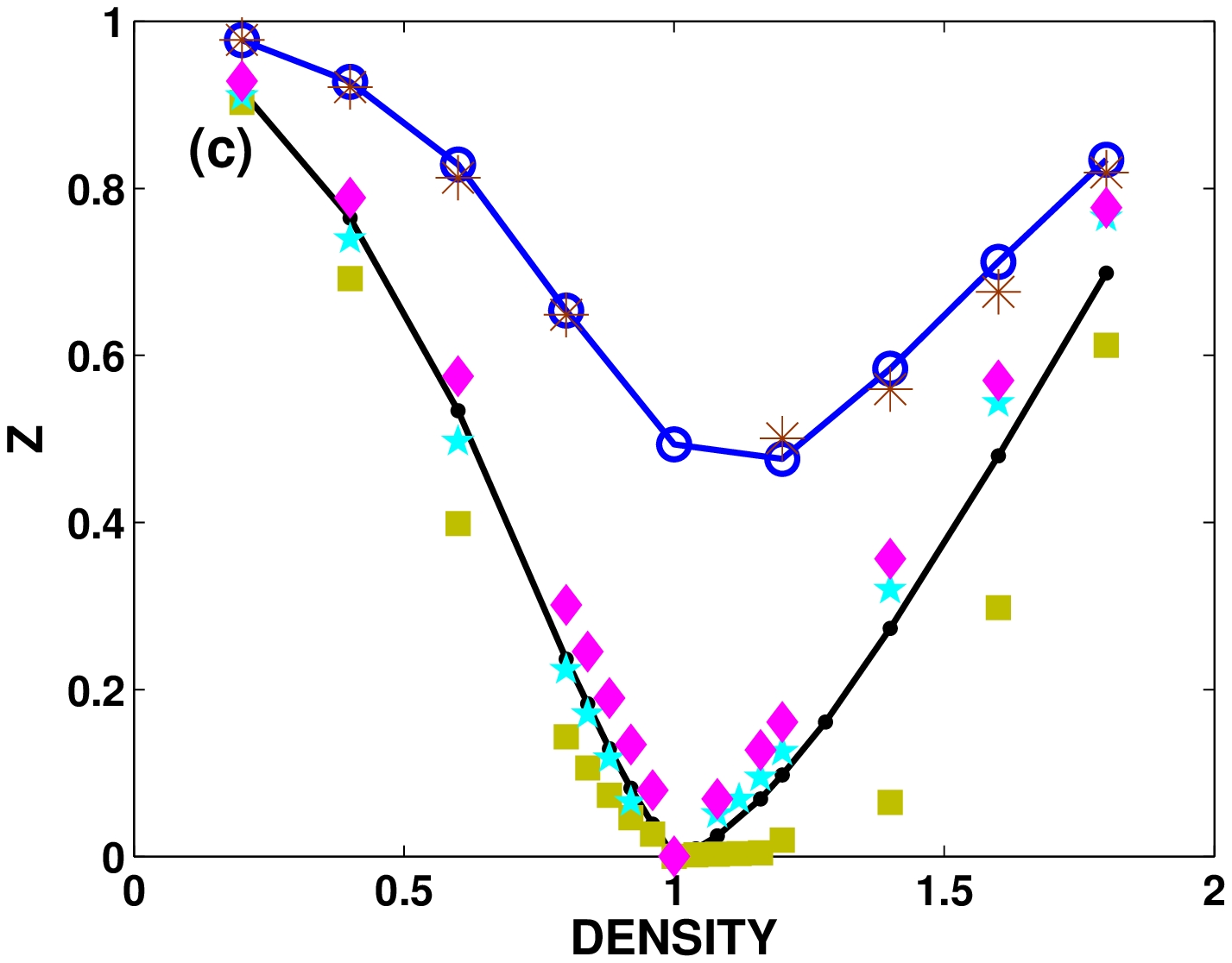}}
\end{center}
\caption{(Color online) Fermi liquid parameters as a function of density. Zero frequency results are obtained with the same extrapolation method as in Fig. 2. (a) Check of Luttinger's theorem. The effective chemical potential $\tilde{\mu}= \mu - \Sigma'(0)$ should equal the non-interacting value, shown in red, when the theorem is satisfied. For $U=8t$, the brown asterisks ($\ast $) obtained with IPT $n=n_0$ satisfy the theorem. For $U=32t$ results for three different methods are shown: in kaki ($\square$) for IPT $n=n_0$, in cyan ($\star$) for IPT $D_{naive}$ and in magenta ($\lozenge$) for IPT $\langle D\rangle_{CTQMC}$. (b) $\Sigma''(0)$ is plotted for $U=32t$ in magenta ($\lozenge$) for IPT $\langle D\rangle_{CTQMC}$ as above, and compared with the CTQMC results shown previously in Fig.\ref{fig:Fermi_CTQMC} (black dots joined by a line). (c) Quasiparticle spectral weight $Z$ computed for different methods and displayed with the same symbols as in (a). We compare with the CTQMC results of Fig. 2, namely blue symbols ($\circ $) with line for $U=8t$ and black symbols (.) with line for $U=32t$. The results for IPT $n=n_0$ at $U=32t$ are un-physical since they predict an insulator away from half-filling.}
\label{fig:Fermi_IPT}
\end{figure}

The results for the low-temperature extrapolations of $\tilde{\mu}= \mu - \Sigma'(0)$ and $Z$ for $U=8t$ and $U=32t$ are shown in Fig.\ \ref{fig:Fermi_IPT}. Below the Mott transition, $U=8t$, the $n=n_0$ results (brown ($\ast $)) are shown. One can detect only a very small difference with the solid red line. Luttinger's theorem is thus essentially satisfied.

On the other hand, IPT-$n_0$ for $U=32t$ (kaki ($\square$)) gives non-physical results not only\cite{Potthoff:1997} for $n>1$ but, quite generally, close to half-filling. Not only is Luttinger's theorem
strongly violated, but for a large range of densities, $n>1$, $\tilde{\mu}$ is outside the band. Many properties of the Fermi liquid are proportional to functions of the non-interacting system evaluated at $\tilde{\mu}$. But these functions are zero outside the band and so if $\tilde{\mu}$ is outside the band we obtain zero. For example, this would predict an insulator away from half-filling. The situation is not better for $Z$, especially around half-filling where it vanishes for a finite range of densities when $n>1$. This demonstrates that at low $T$, when $U$ is large, IPT-$n_0$ in its simplest form cannot be applied. We must thus search for a new condition to explore this region of parameter space.

\section{IPT Double occupancy: IPT-$D$}\label{Sec:Double}
Imposing exact results such as sum-rules, whenever possible, is desirable for any physical theory. Whereas the condition $n=n_0$ is not required by any fundamental principle, the self-energy must always obey
\begin{equation}\label{D_occ_sigG}
    D = \d{T}{U}\sum_n\text{e}^{i\omega_n0^+}\Sigma (i\omega_n)G(i\omega_n).
\end{equation}
where $D=<n_{\uparrow}n_{\downarrow}>$ is double occupancy. Enforcing this consistency condition between single-particle properties, such as $\Sigma$ and $G$, and a two-particle property, $D$, has been successful in other approaches, such as the Two-Particle-Self-Consistent theory.\cite{Vilk:1997,TremblayMancini:2011}. In the regime of interest here, strong coupling, $D$ can be accurately estimated and is only very weakly dependent on temperature, as discussed in the following subsection. There, we assess the accuracy of the approach.

\subsection{Exact and naive values of double occupancy at strong coupling}
\begin{figure}[tbp]
\includegraphics[width=0.9\linewidth]{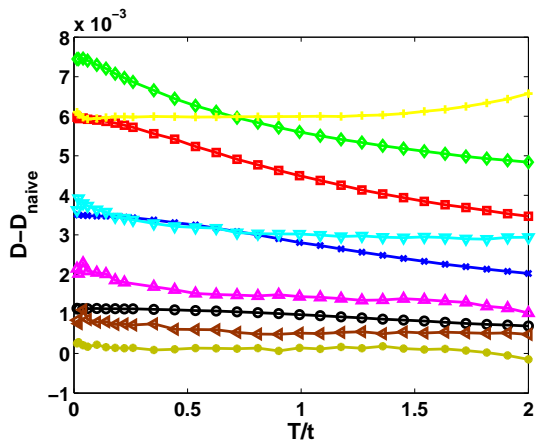}
\caption{(Color online) CTQMC results at $U=32t$ for double occupancy $D-D_{naive}$ plotted as a function of temperature. We define $D_{naive} = 0$ for fillings $n<1$ and $D_{naive} = n-1$ for $n>1$. The various densities are represented by different symbols: $n=0.2$ (black ($\circ $)), $0.4$ (blue ($%
\times $)), $0.6$ (red ($\square $)), $0.8$ (green ($\lozenge $)), $1.0$ (yellow (+)), $1.2$
(cyan ($\triangledown $)), $1.4$ (magenta ($\vartriangle $)), $1.6$ (brown ($\vartriangleleft $)) and $1.8$ (kaki ($\star $)). The largest deviations from the naive value, occurring close to $n=1$, are less than $10^{-2}$ in absolute value.} \label{fig:D_occ}
\end{figure}

\begin{figure}[tbp]
\begin{center}
\includegraphics[width=0.9\linewidth]{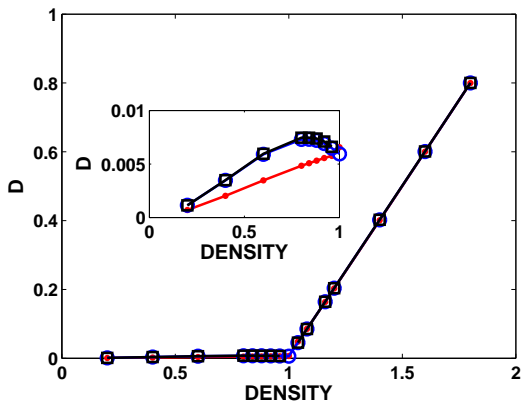}
\end{center}
\caption{(Color online) Double occupancy $D$ as a function of density obtained from CTQMC for $U=32t$ for three temperatures: $\beta = 25/t$ (black ($\square $)), $\beta = 10/t$ (blue ($\circ$)) and $\beta = 0.5/t$ (red ($\cdot$)). On this scale, the naive value of $D$ is very accurate. The inset is a zoom for densities $n \leq 1$.}
\label{fig:D_occ_n}
\end{figure}

For very large $U$, it is easy to guess that $D$ should depend only very weakly on temperature. In addition, the value of $D$ can be estimated quite accurately. Indeed if $n<1$ there are unoccupied sites in the lattice and since $U$ is large, $D$ should naively be zero. For $n>1$, $D$ is necessarily non-zero. If we start from the half-filled, Mott insulator, with one electron per site and add electrons, they must go to a site which is already occupied. Thus $D$ should simply equal the excess number of electrons measured from half-filling, $D = n-1$. These estimates are called $D_{naive}$. In reality for $n<1$ there are corrections of order $t/U$ to double occupancy, giving rise to exchange interaction, and $D$ is slightly larger than zero. For analogous reasons, $D$ for $n>1$ will always be a bit larger than $n-1$ i.e. $D= (n-1) + \delta D$. This is obvious for models with particle-hole symmetry, but it will be true as well, even in the absence of particle-hole symmetry.

We can verify our estimates with the CTQMC results for $U=32t$. Within CTQMC, $D$ is calculated directly on the impurity by the Monte-Carlo sampling. Fig.\ref{fig:D_occ} displays $D-D_{naive}$ as a function of temperature for different densities. We see that $D-D_{naive}$ is small and that the $T$ dependence is on the third significant digits. So, for all practical purposes, we can assume $D$ to be independent of $T$ although it differs from $D_{naive}$. The values of $D$ obtained are very close to the naive expectation but always slightly larger. In Fig.\ref{fig:D_occ_n} we show the double occupancy $D$ as a function of the density for three different temperature from $T = 0.04t$ to $T = 2t$. This figure confirms again that even if we have some dependence on $T$, it is quite small and the result is a fairly simple function of the density. Very similar results have been obtained in Ref. [\onlinecite{DeLeo:2011}] in the case of a 3d simple cubic lattice. We thus have a rather simple constraint that we can take into account in IPT to fix all parameters. We call this approach IPT-$D$. In the next subsection, we will assess the accuracy of this approach and verify how the results are modified when the exact value of $D$ is used instead of the naive one.

\subsection{Accuracy of Fermi liquid parameters IPT-$D$}
We first set $D$ to its naive values, i.e zero for $n\leq 1$ and $n-1$ for $n>1$. The results for $\tilde{\mu}$ and $Z$ are shown in Fig.~\ref{fig:Fermi_IPT} as cyan stars. Clearly, the value obtained for $\tilde{\mu}$ is in much better agreement with Luttinger's theorem than in the case IPT-n$_0$, although it is still incorrect for densities near half-filling. The biggest improvement is that we avoid values of $\tilde{\mu}$ outside the band. Furthermore, in the case of $Z$, IPT-$D$ is close to the CTMQC results while for IPT-n$_0$ it is quite far\cite{Potthoff:1997} from the correct result, leading in particular to an un-physical insulator over a finite range of densities for $n>1$.

As we now show, one can improve the results further by using an accurate value of $D$. That value can be obtained from a number of methods, in particular CTQMC. It can be computed quite accurately and does not require a large number of Matsubara frequencies. Even if CTQMC is available, it may be desirable to use IPT-$D$ because the calculation can either be done directly in real frequency or analytically continued from Matsubara frequencies using simple methods such as Pad\'e approximants\cite{Vidberg:1977}, whereas with CTQMC, Maximum entropy\cite{Jarrell:1992} is necessary.  In addition, since $D$ has negligible temperature dependence in strong coupling, only one value of $D$ may be sufficient with IPT-$D$ to compute other quantities for a wide range of temperatures.

Since $D$ is not completely $T$ independent, we use an average called $\langle D\rangle_{CTQMC}$ calculated between $\beta = 75/t$ and $\beta = 0.5/t$ for the purpose of comparison with the naive approach. It is calculated from the arithmetic mean of the numerical values of $D(T)$. Note that the values of $D(T)$ for each $n$ are taken as the arithmetic mean of the last four DMFT iterations. The results are shown in Fig.~\ref{fig:Fermi_IPT} as magenta lozenges. We see that the results for $\tilde{\mu}$ are in excellent agreement with Luttinger's theorem except very close to half-filling where it deviates, but not too much. Surprisingly, $Z$ is not as accurate as that obtained from the naive estimate of $D$. But if we look at the results for densities between 0.8 and 1.2, the difference between $Z_{\langle D\rangle_{CTQMC}}$ and $Z_{CTQMC}$ is small and constant and $Z_{\langle D\rangle_{CTQMC}}$ correctly extrapolates to zero at half-filling. A small difference in $D$ can have a quantitative impact on $Z$, without affecting qualitative trends. For example for $n=0.84$, the $D$ given by CTQMC is $D=0.00619$ instead of the naive $D=0$, whereas for $n = 1.16$, CTQMC gives $D=0.1637$ instead of $D=0.16$. We could imagine that because these quantities are at low $T$, it would be better to take a $D$ that is in the low temperature range. If we do this, we obtain a $\langle D\rangle_{CTQMC}$ slightly larger. We then find that $\tilde{\mu}$ is not really affected while $Z$ is a little bit worse than that obtained from the average $D$ over the larger $T$ range. Some tuning of $D$ would allow us to get a best possible set of $\tilde{\mu}$ and $Z$, but that is clearly not the purpose of the exercise. Finally, for $\Sigma''(0)$ Fig.~\ref{fig:Fermi_IPT}-(b) shows that IPT-$D$ is qualitatively correct while being always smaller than CTQMC.

\subsection{Accuracy of the Density of States and Chemical Potential}
\begin{figure}[tbp]
\centering
\subfigure{\includegraphics[width=0.9\linewidth]{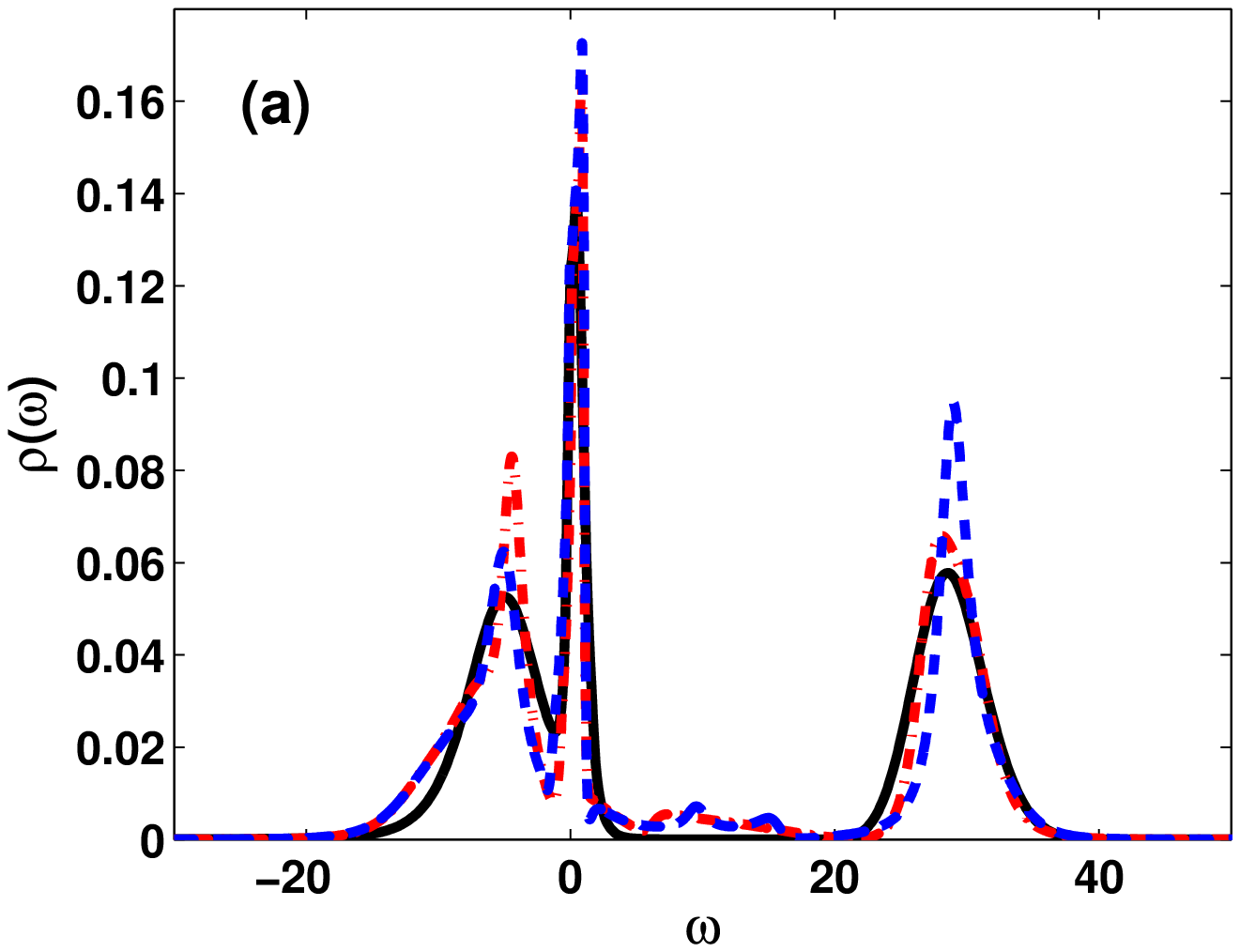}}
\subfigure{\includegraphics[width=0.9\linewidth]{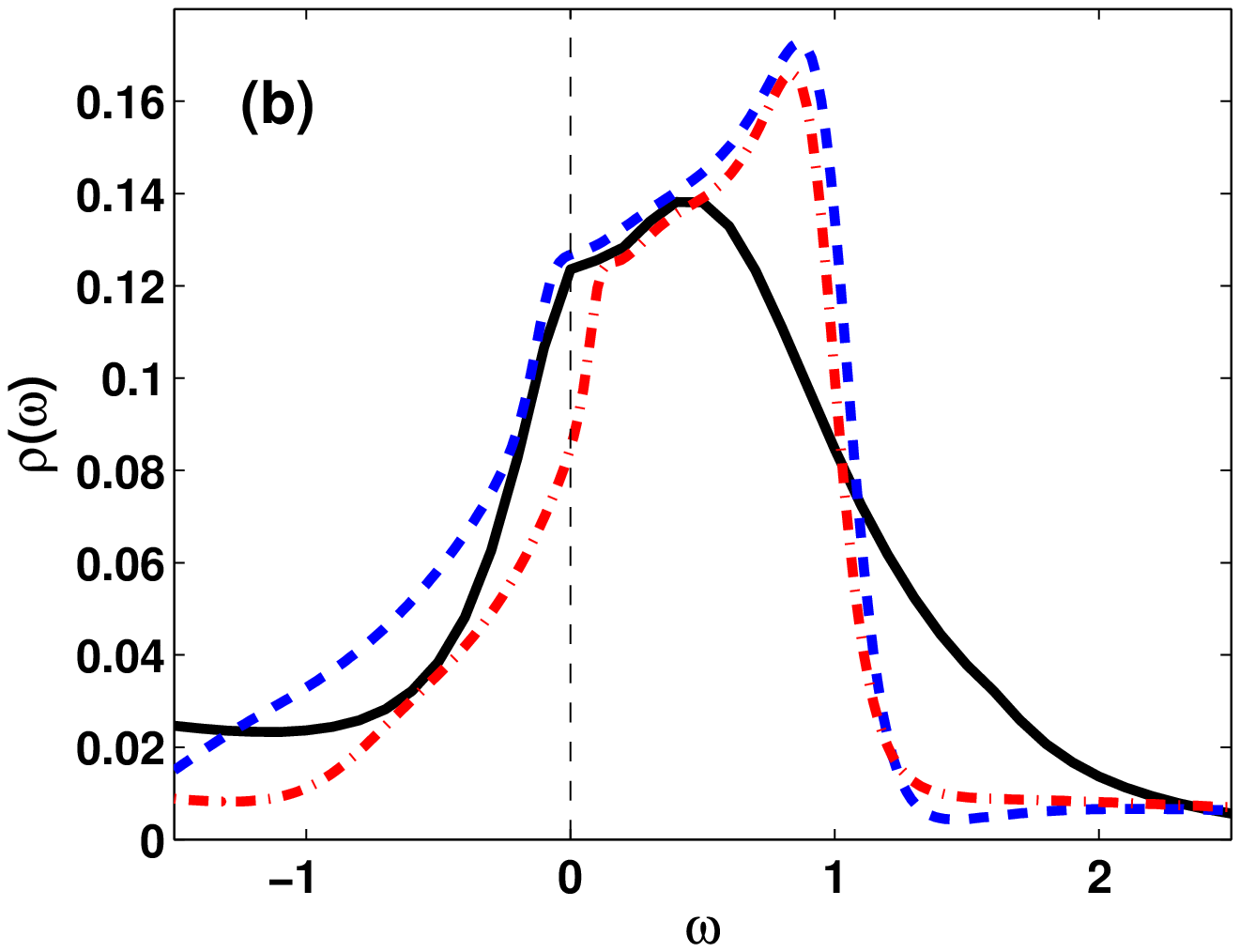} }
\caption{(Color online) Density of states for $n = 0.84$, $\beta t = 25$ and $U=32t$ obtained with three methods: black (solid line) with CTQMC maxent, blue (- -) with IPT-$\langle D\rangle_{CTQMC}$, and red (-.) with IPT-$D_{naive}$. (b) is a zoom of (a) around $\omega = 0$. The value at zero frequency is improved when a more accurate value of $D$ is used in IPT. } \label{fig:DOS}
\end{figure}

 It is instructive to look at the Density of States obtained from IPT with fixed $D$ and Pad\'e analytical continuation\cite{Vidberg:1977}. We show $n = 0.84$ and $\beta = 25/t$ as typical values in Fig.\ref{fig:DOS}. We compare to the CTQMC values obtained from Maximum Entropy analytical continuation of $G(i\omega_n)$. That Green's function is an average over several converged DMFT iterations. The Maximum Entropy implementation that was used here is somewhat crude and thus we must not really focus on the details. The CTQMC results have different errors at different scale, i.e very precise at low $\omega_n$, fluctuating at intermediate $\omega_n$ while at large $\omega_n$ the results are analytical. Hence, we choose the weight of the entropy term based on heuristic considerations, depending on the real-frequency range we are interested in.

 As was noted previously\cite{Kajueter:1996,PhDKajueter}, in IPT there are states in the Mott gap at finite frequency, but their weight is small compared to the states everywhere else, namely near zero frequency and in the lower and upper Hubbard band. Overall, IPT with fixed $D$ compares well with CTQMC, but, at low temperature, what really matters is the region near $\omega = 0$. We thus zoom on this region in Fig.\ \ref{fig:DOS}-(b). There, we see that in the vicinity of $\omega = 0$, when $D=\langle D\rangle_{CTQMC}$, we are quite close to the CTQMC values. When we use the naive $D$, the quasi-particle peak is shifted a little bit to the right and so is this why the low $T$ results are different even if the shape and values of the peaks are similar.

\begin{figure}
\begin{center}
\mbox{\includegraphics[width=0.9\linewidth]{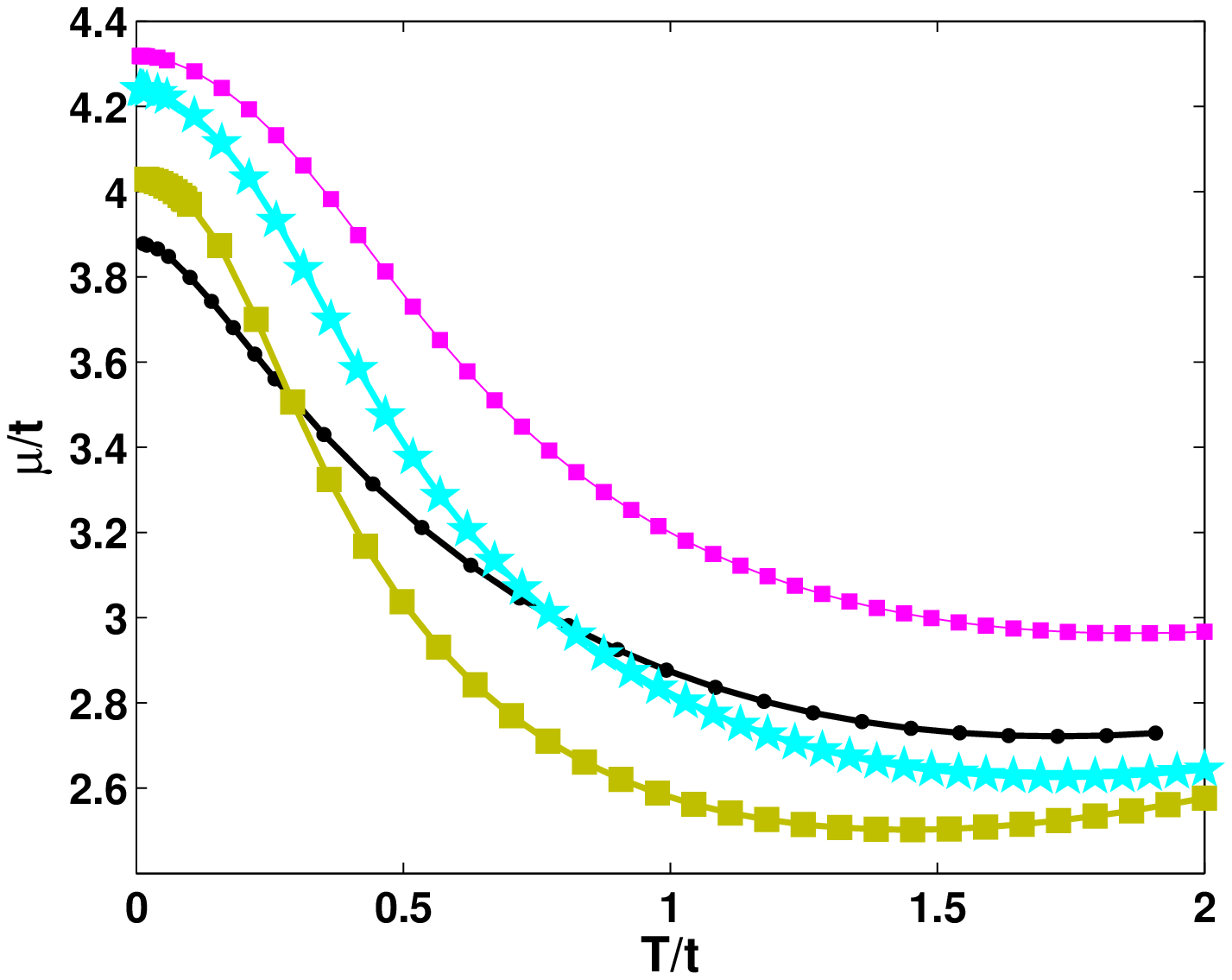}}
\mbox{\includegraphics[width=0.9\linewidth]{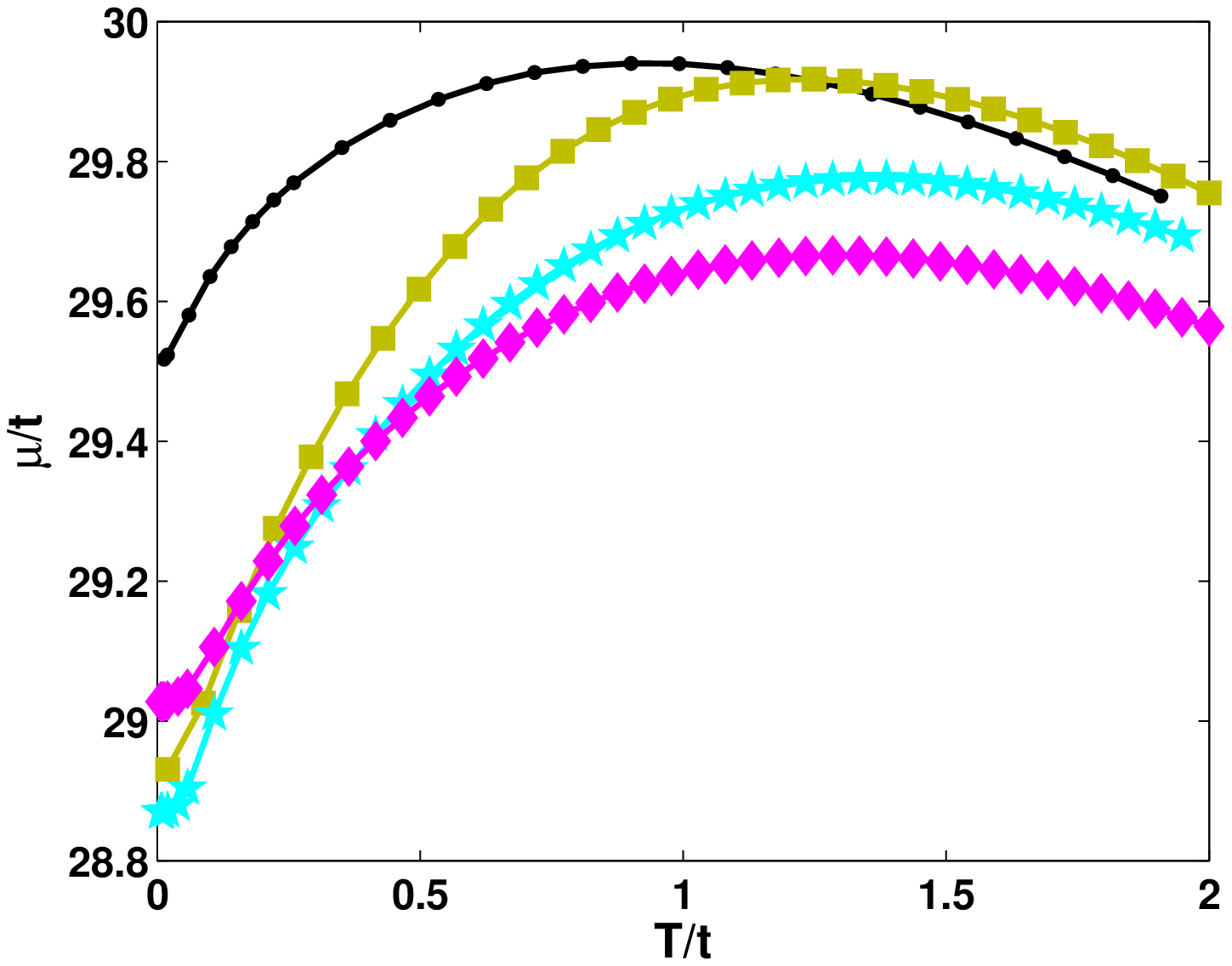}}
\mbox{\includegraphics[width=0.9\linewidth]{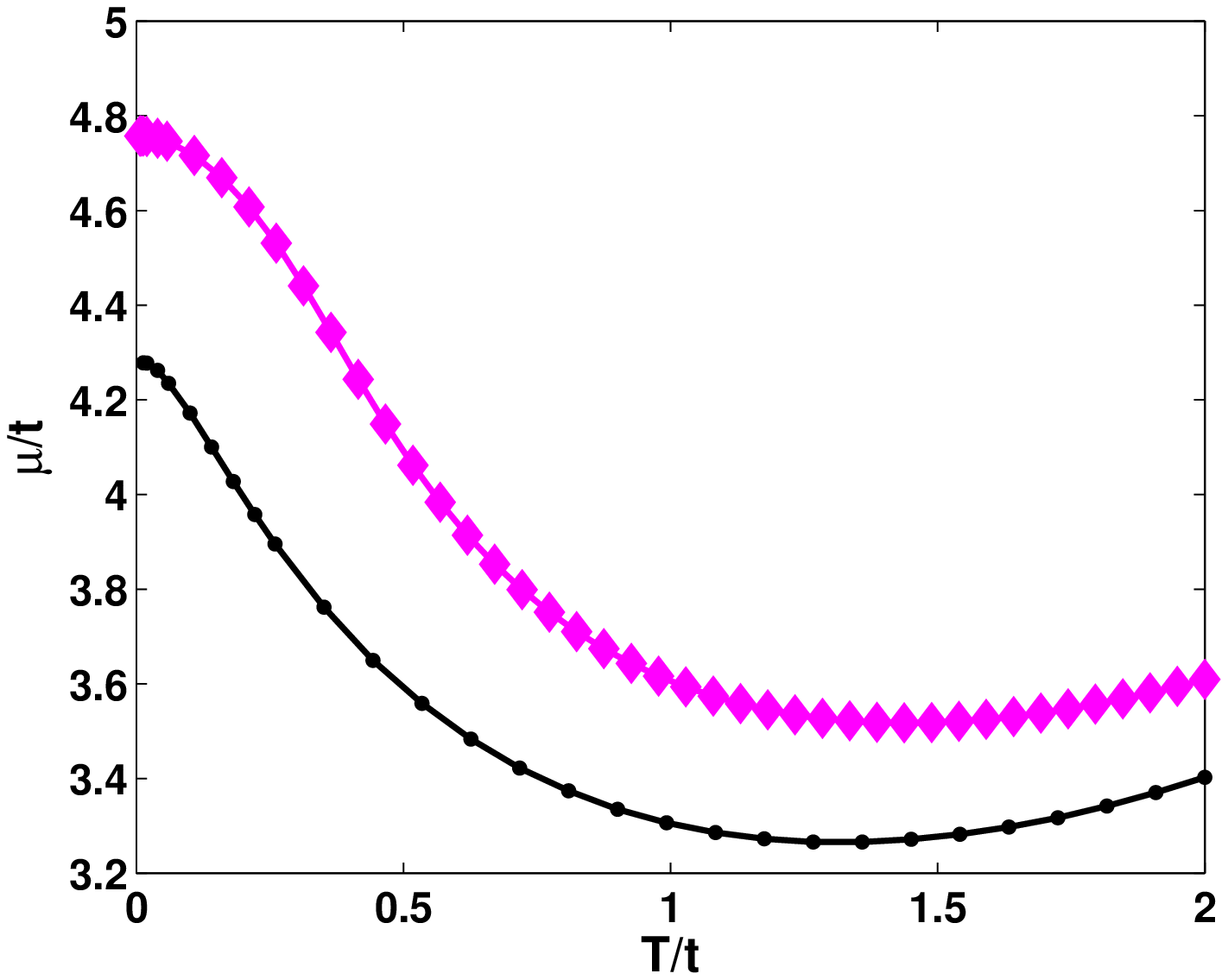}}
\end{center}
\caption{(Color online) Chemical potential as a function of temperature for different IPT approximations, compared with the reference CTQMC calculations as black dots with line obtained for $U=32t$ and different densities: (a) $n=0.80$, (b) $n=1.2$ and (c) $n=0.84$. The three different IPT approximations are given in kaki ($\square$) for IPT-n$_0$, in cyan ($\star$) for IPT-$D_{naive}$,  and in magenta ($\lozenge$) for IPT-$\langle D\rangle_{CTQMC}$. The latter approximation in magenta ($\lozenge$) is best, having a more or less doping and temperature independent offset $\delta\mu/t\sim 0.5$ when compared with the reference CTQMC in black.}
\label{fig:mu_T}
\end{figure}

We also compare the results for an integrated quantity, $\mu (T)$, that is obtained in general by solving $n = 2\int f(\omega)\rho (\omega)d\omega$ with $f(\omega)$ the Fermi function and $\rho(\omega)$ the density of states. In our case, this is a byproduct of the DMFT calculation. No analytical continuation is involved. In Fig.~\ref{fig:mu_T} we show the results for three densities. In Fig.~\ref{fig:mu_T}(a) and (b) we note that the numerical values obtained with the different methods differ by at most about 10\%. The best results are for IPT-$\langle D\rangle_{CTQMC}$ since the curves are qualitatively very similar to the CTQMC ones with a derivative quantitatively quite close for all $T$. The absolute difference between IPT-$\langle D\rangle_{CTQMC}$ and CTQMC is almost doping and temperature independent. The derivative with respect to temperature for both IPT-$n_0$ and IPT-$D_{naive}$ is not as good. At high enough temperature, all methods give similar results. Fig.~\ref{fig:mu_T}(c) shows the result closer to half-filling, comparing CTQMC with IPT-$\langle D\rangle_{CTQMC}$, the best IPT method. As already discussed, IPT-$n_0$ gives un-physical results in the vicinity of half-filling.

\subsection{Energy and Specific Heat}
In this section we compare internal energy and specific heat in IPT-$D$ with CTQMC. Within CTQMC, energy can be calculated quite accurately with a reasonable number of Matsubara frequencies. Indeed, it was shown by Haule\cite{haule:2007i} that the kinetic energy $\langle K\rangle$ is proportional to the average perturbation order $\langle k \rangle$ for a given set of parameters. As already discussed, the double occupancy is calculated directly by CTQMC and thus the total energy is given by
\begin{equation}\label{energy_CTQMC}
    E_{CTQMC}(T) = -T\langle k \rangle + UD.
\end{equation}
In general, for a many-body system, the energy is given by the thermal average, in the grand-canonical ensemble, of the Hamiltonian. For the Hubbard model we may write
\begin{equation}\label{energy_gen}
\begin{split}
    E(T) &= \frac{1}{N}\sum_{k,\sigma}\varepsilon_k \langle d^{\dagger}_{k,\sigma}d_{k,\sigma}\rangle +  \frac{U}{N}\sum_{i}\langle n_{i\uparrow }n_{i\downarrow }\rangle\\
    &= \mu n + \frac{1}{N}\sum_k\varepsilon_k G_k(\tau = 0^-) - \frac{1}{N}\sum_k\d{\pd G_k(\tau)}{\pd\tau}\Big|_{\tau = 0^-}\\
    &= \d{1}{\beta}\frac{1}{N}\sum_{k,n}\text{e}^{-i\omega_n0^-}\left[i\omega_n + \varepsilon_k + \mu\right]G_k(i\omega_n).
\end{split}
\end{equation}
For IPT-$D$, we use directly the imaginary time expression. Since we generally compute Green's functions only for positive imaginary time, we use the equivalent expression
\begin{equation}
\begin{split}
E(T) &= \mu n + Un - \mu + \frac{1}{N}\sum_k\varepsilon_k G_k(\tau = 0^+)\\
 &- \frac{1}{N}\sum_k\d{\pd G_k(\tau)}{\pd\tau}\Big|_{\tau = 0^+},
\end{split}
\end{equation}
where we used that for the FCC lattice $\sum_k\varepsilon_k = 0$. Once the energy is calculated, the specific heat $C_n$ is given by $C_n = \d{dE(T)}{dT}$.

\begin{figure}[H]
\subfigure{\includegraphics[width=0.9\linewidth]{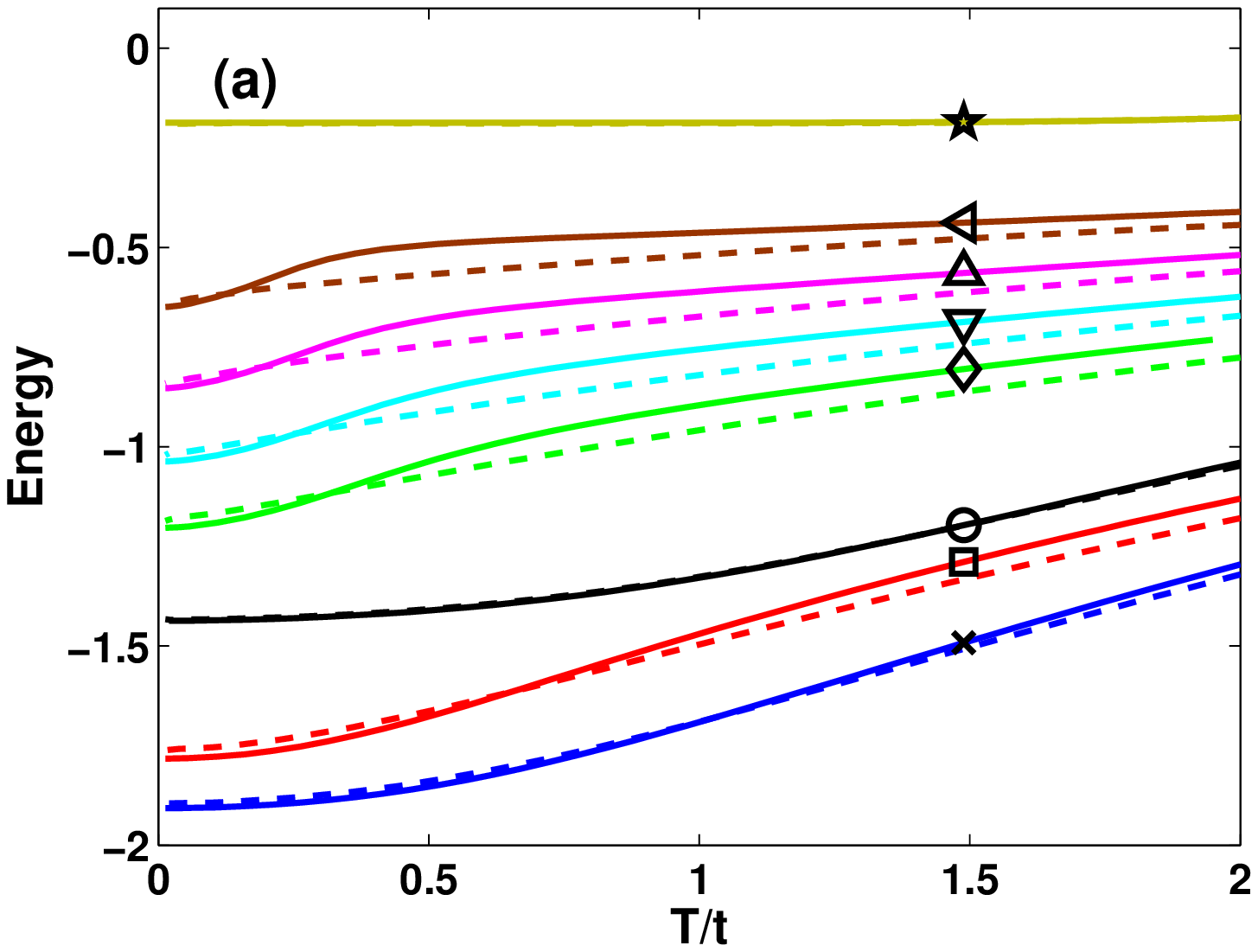} }
\subfigure{\includegraphics[width=0.9\linewidth]{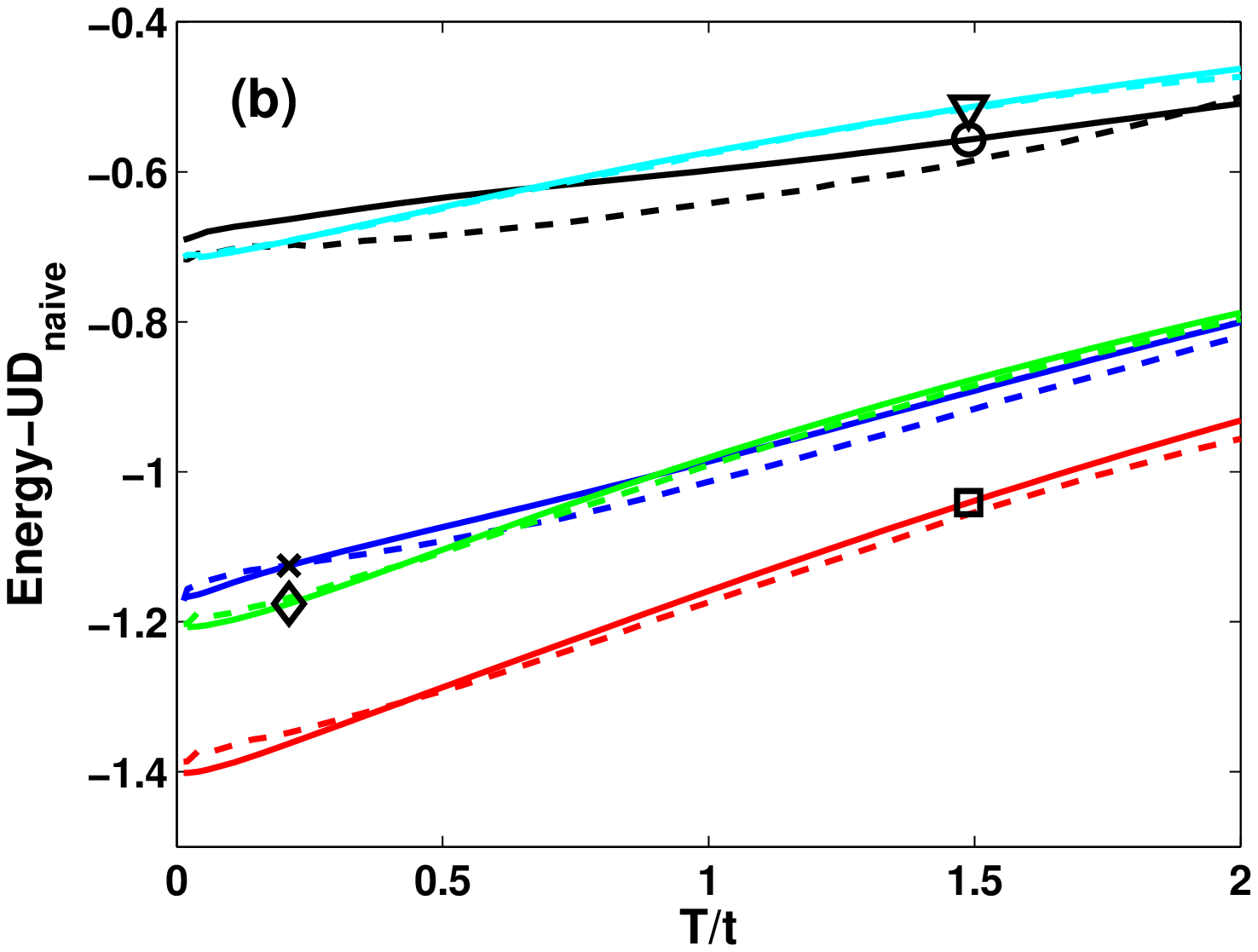} }
\centering
\mbox{\subfigure{\includegraphics[width=0.45\linewidth]{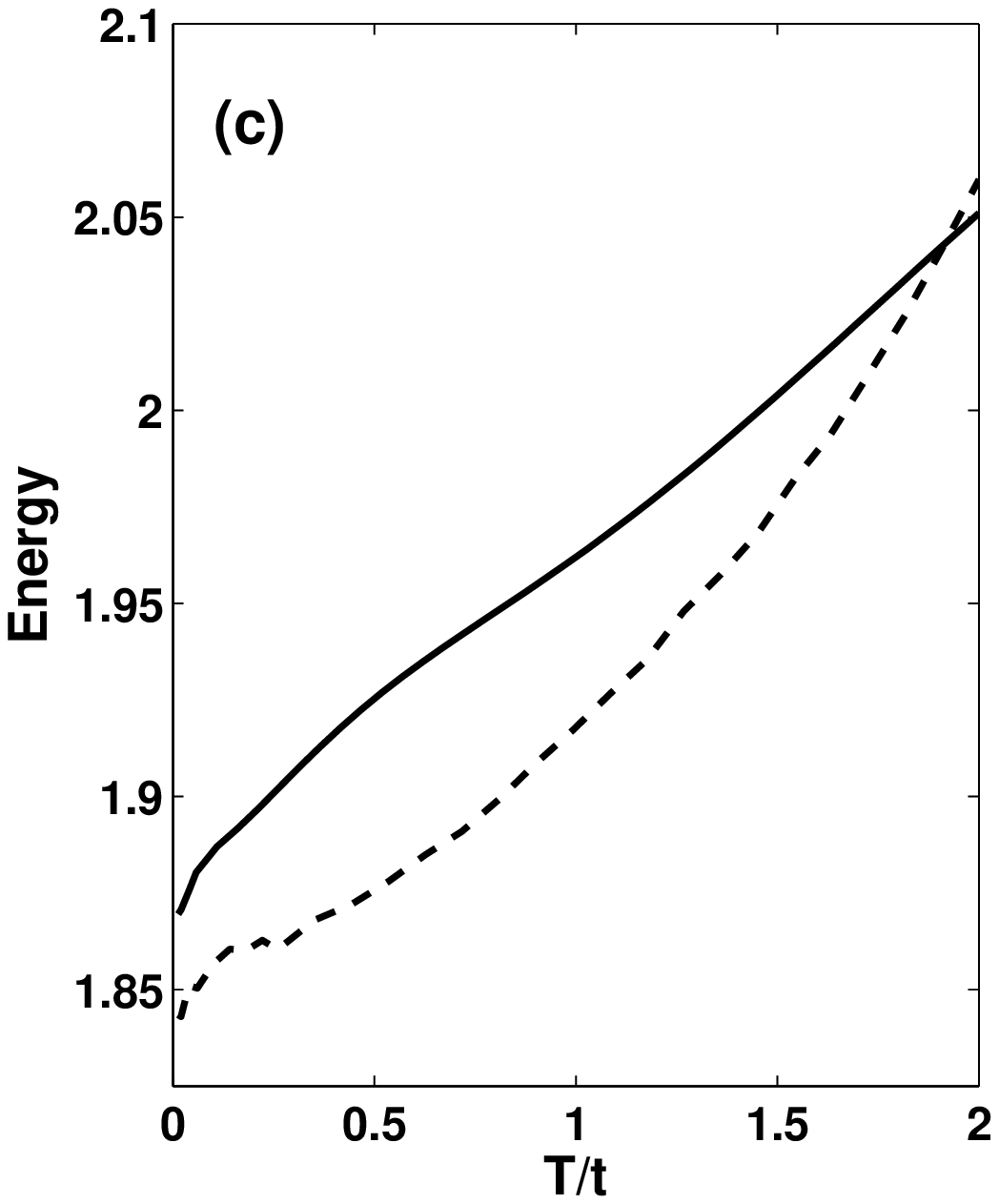}}\quad
\subfigure{\includegraphics[width=0.45\linewidth]{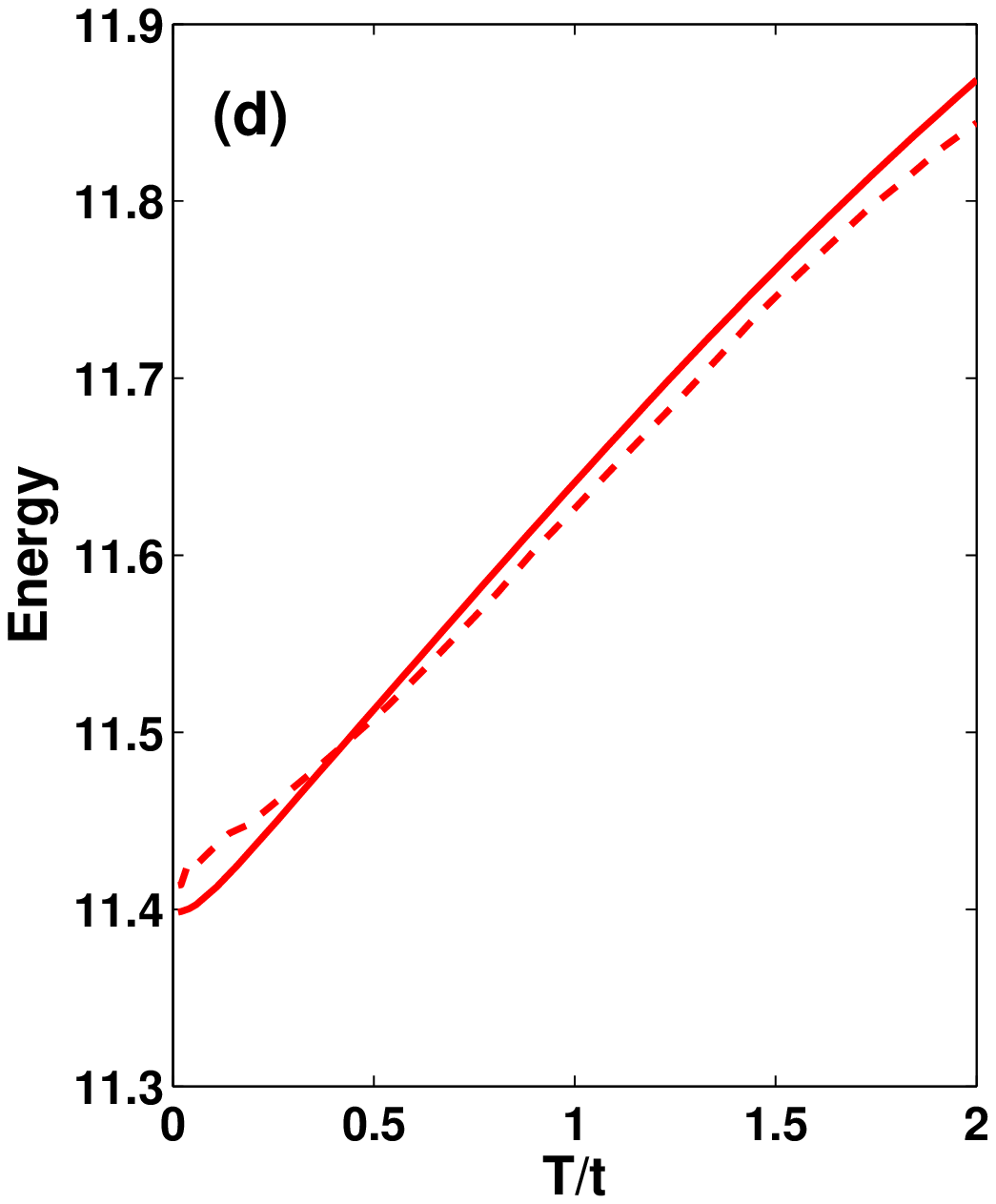} }}
\caption{(Color online) Energy as a function of temperature obtained from IPT-$D$ (solid lines) and CTQMC (dashed lines) for $U=32t$. (a) Densities equal to, or below half-filling $n=0.2$ (black ($\circ $)), $0.4$ (blue ($\times $)), $0.6$ (red ($\square $)), $0.8$ (green ($\lozenge $)), $0.84$
(cyan ($\triangledown $)), $0.88$ (magenta ($\vartriangle $)), $0.92$ (brown ($\vartriangleleft $)) and $1.0$ (kaki ($\star $)). For densities above half-filling, displayed in (b), $n=1.08$ (black ($\circ $)), $1.2$ (blue ($\times $)), $1.4$ (red ($\square $)), $1.6$ (green ($\lozenge $)), $1.8$
(cyan ($\triangledown $)), the quantity $UD_{naive}$ is subtracted from the energy to allow the results to fit on the same scale. (c) is a zoom for $n = 1.08$ and (d) a zoom for $n=1.4$.} \label{fig:Energy}
\end{figure}
\begin{figure}[H]
\subfigure{\includegraphics[width=0.9\linewidth]{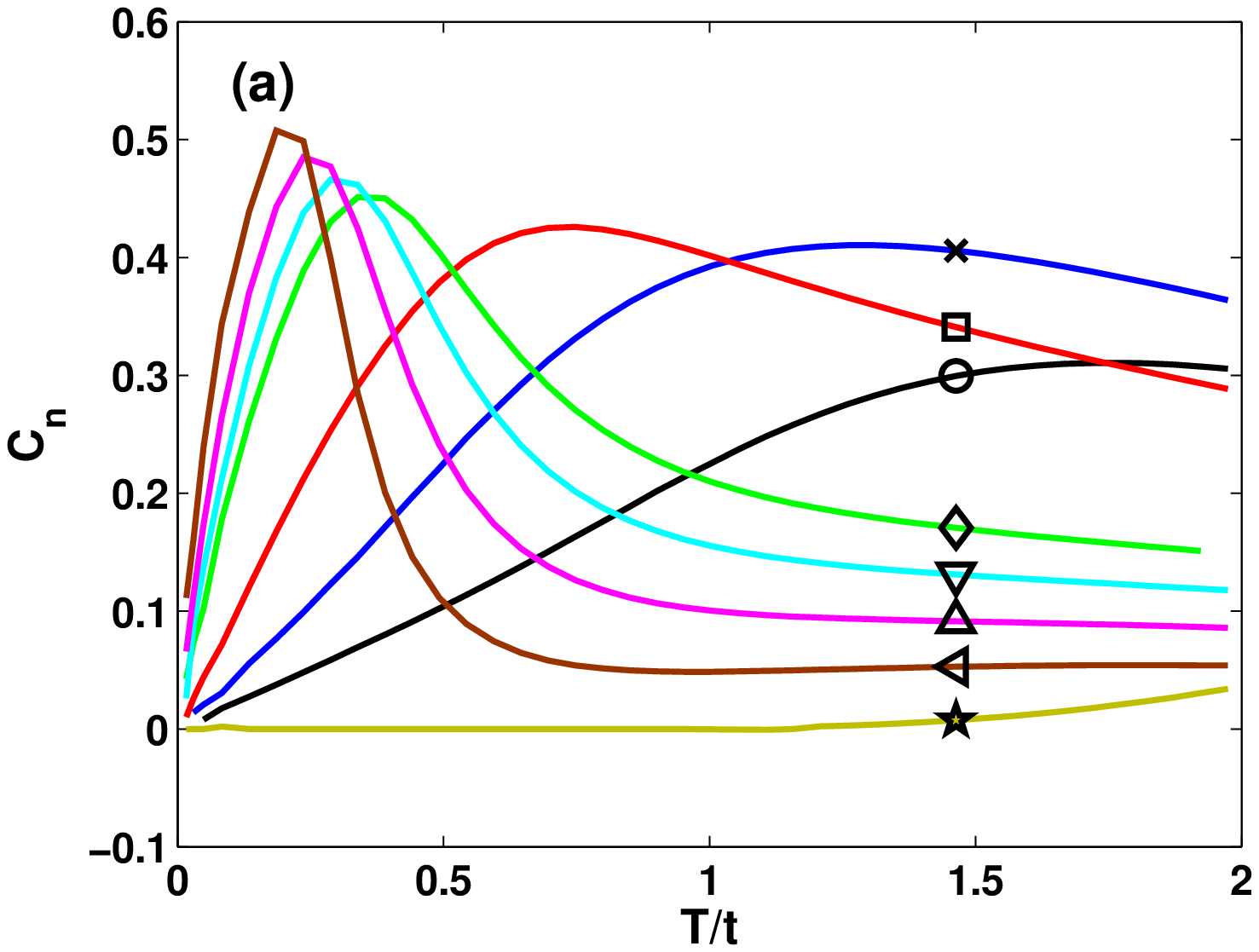} }
\subfigure{\includegraphics[width=0.9\linewidth]{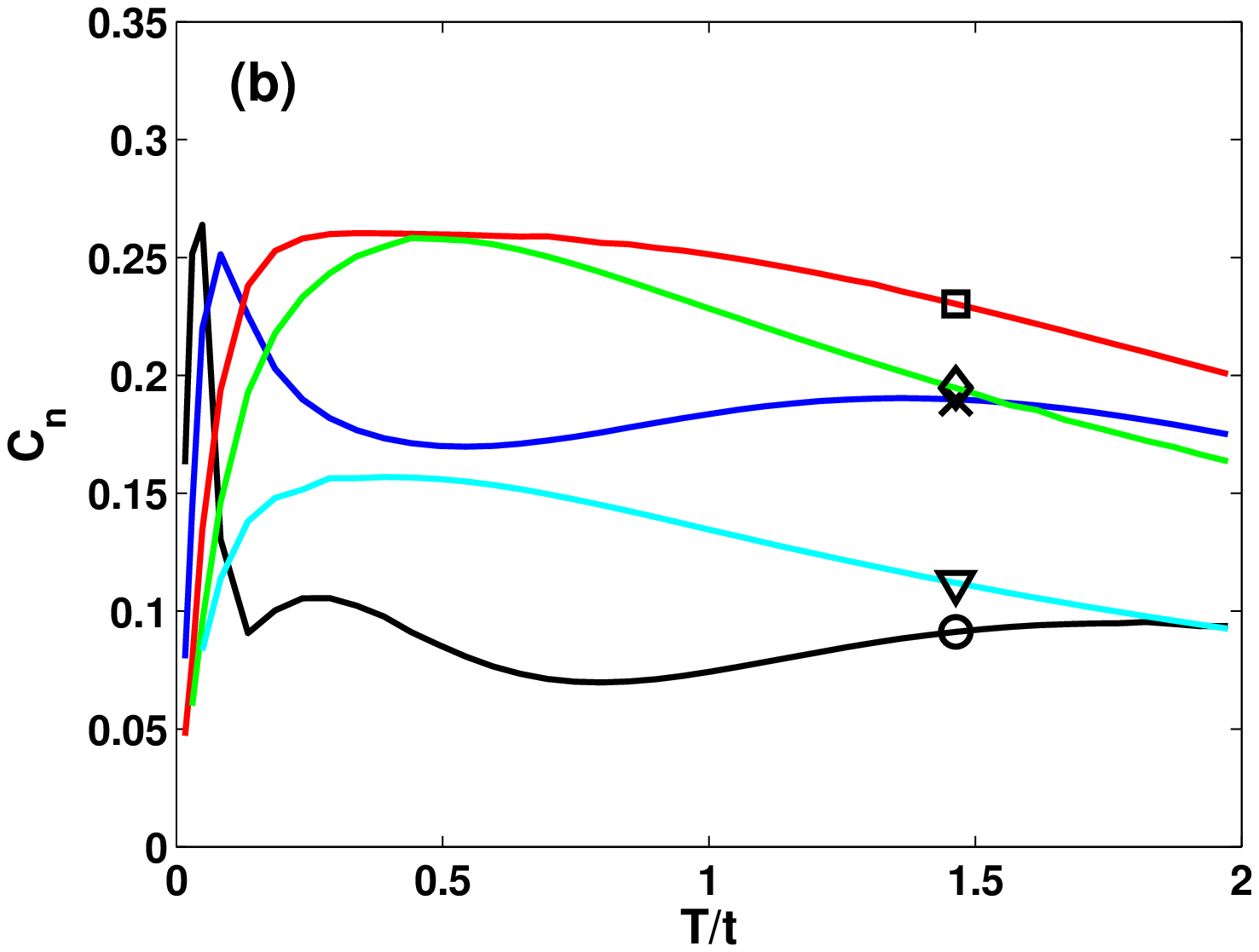} }
\subfigure{\includegraphics[width=0.9\linewidth]{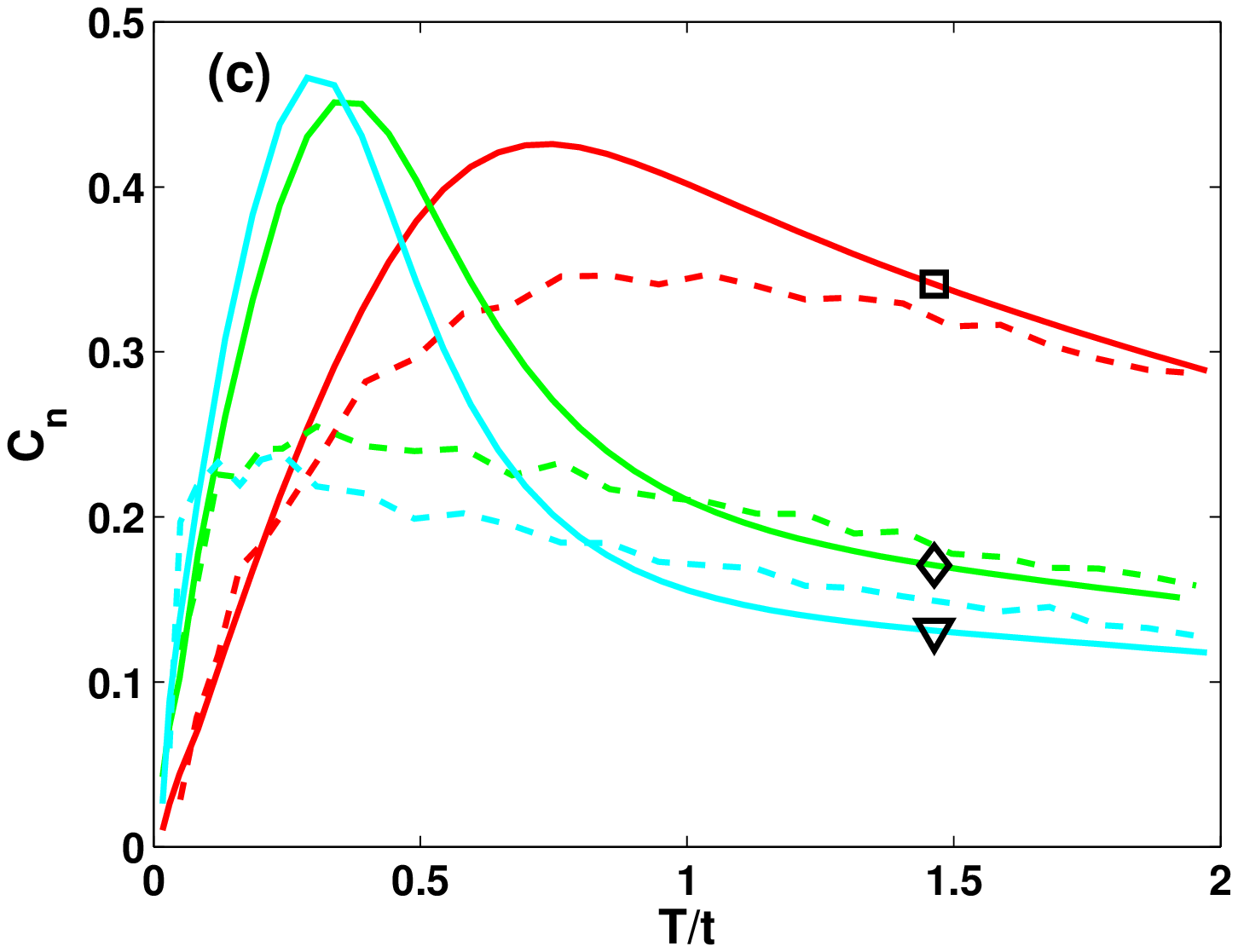} }
\caption{(Color online) Specific heat at constant filling as a function of temperature for $U=32t$. (a) Results from IPT-$D$ (solid line) for densities below half-filling: $n=0.2$ (black ($\circ $)), $0.4$ (blue ($\times $)), $0.6$ (red ($\square $)), $0.8$ (green ($\lozenge $)), $0.84$(cyan ($\triangledown $)), $0.88$ (magenta ($\vartriangle $)), $0.92$ (brown ($\vartriangleleft $)) and $1.0$ (kaki ($\star $)), (b) Specific heat from IPT-$D$ (solid line) for densities above half-filling $n=1.08$ (black ($\circ $)), $1.2$ (blue ($\times $)), $1.4$ (red ($\square $)), $1.6$ (green ($\lozenge $)), $1.8$
(cyan ($\triangledown $)), (c) Comparison between IPT-$D$ as solid lines and CTQMC as dashed lines for $n = 0.6$ (red ($\square $)), $0.8$ (green ($\lozenge $)) and $0.84$(cyan ($\triangledown $)). The peak positions and shape coincide, even though the absolute values differ. In this case, since CTQMC values comme from differentiation of Monte Carlo data, there is a rather large uncertainty, especially for peaks.} \label{fig:Cv}
\end{figure}

To obtain values for all temperatures, we have considered the fixed $D$ as the one given by the average over $0.5 \leq \beta \leq 75/t$ of the CTQMC results. This is what we called IPT-$\langle D\rangle_{CTQMC}$ previously. We use IPT-$D$ for brevity. We present the comparison in Fig.\ref{fig:Energy} for different densities. Note the change in color code (see the legend). In Fig.\ref{fig:Energy}-(a) we display results for densities $n\leq 1$ while in Fig.\ref{fig:Energy}-(b), the density is above half-filling $n > 1$. The energy scale for the latter case is shifted by a filling dependent quantity $UD_{naive}$ so that the various fillings can be displayed on the same scale. In Fig.\ref{fig:Energy}-(c) and (d) we zoom on $n = 1.08$ and $n = 1.4$ respectively without energy shift to emphasize the differences between IPT-$D$ and CTQMC. Comparing those differences in Fig.\ref{fig:Energy}-(c) with those in Fig.\ref{fig:Energy}-(d), we see that the further we are from half-filling, the better the agreement.

In Fig.\ref{fig:Cv} we plot the specific heat as predicted by IPT-$D$. Once again we consider separately $n\leq 1$ in Fig.\ref{fig:Cv}-(a) and $n > 1$ in Fig.\ref{fig:Cv}-(b). The closer we are to half-filling, the lower the temperature at which the peak that signals the appearance of the Fermi liquid regime appears. Despite the strong coupling, the particle-hole asymmetry is noticeable.

In Fig.\ref{fig:Cv}-(c) we can compare the IPT-$D$ and the CTQMC results for three densities $n=0.60$, $0.80$ and $0.84$. The CTQMC results for the energy are not completely smooth especially at low $T$ because of statistical errors in the data. Thus, if we were to calculate the specific heat $C_n$ for CTQMC, we would, for many of the values of $n$, have to be very careful before calculating the derivative. Here, we just performed a simple derivative on the raw CTQMC data to verify the trend. From the results for the energy, Fig.\ref{fig:Energy}-(a) we already knew that even if the change of curvature in the coherent-incoherent region exists also for CTQMC, it is much smoother than the one obtained from IPT. This is apparent indeed in $C_n$. Nevertheless, the transition temperature for the coherent-incoherent crossover given by the position of the maximum is quite similar for both methods even if, for CTQMC, its exact value is hard to really pinpoint due to the statistical errors in the raw data.

We end this subsection by noting that the inadequacy of IPT-$n_0$ also shows when the energy is calculated. One finds an increase of the energy with decreasing $T$ at low $T$, which is of course non physical since this corresponds to a negative specific heat.

\subsection{DC resistivity}

\begin{figure}[tbp]
\begin{center}
\includegraphics[width=0.9\linewidth]{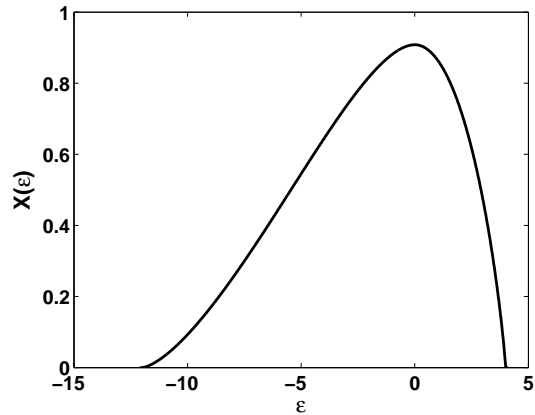}
\end{center}
\caption{Transport function $X(\varepsilon) = \sum_k\left(\d{\pd\varepsilon_k}{\pd k_x}\right)^2\delta(\varepsilon-\varepsilon_k)$, as calculated in appendix~\ref{MC_function}.}
\label{fig:X_func}
\end{figure}
\begin{figure}[tbp]
\begin{center}
\includegraphics[width=0.9\linewidth]{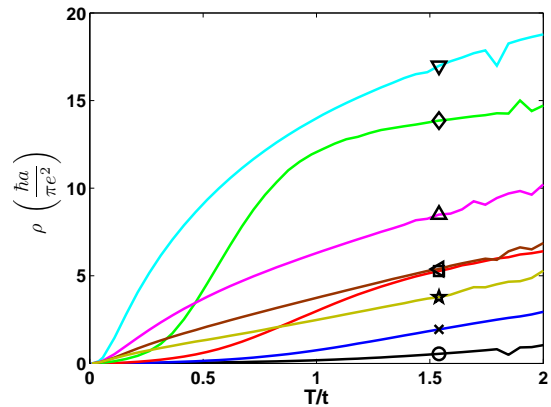}
\end{center}
\caption{(Color online) Resistivity as a function of temperature for $U = 32t$ as calculated by IPT-$D$ for
different values of density: $n=0.2$ (black ($\circ $)), $0.4$ (blue ($%
\times $)), $0.6$ (red ($\square $)), $0.8$ (green ($\lozenge $)), $1.2$
(cyan ($\triangledown $)), $1.4$ (magenta ($\vartriangle $)), $1.6$ (brown ($%
\vartriangleleft $)) and $1.8$ (kaki ($\star $)). The resistivity are largest close to half-filling where they exhibit low coherence temperatures.}
\label{fig:resis}
\end{figure}

Analytical continuation of response functions obtained with CTQMC is in general very difficult. With IPT, one can calculate response functions by first analytically continuing the self-energy with Pad? approximants and using the real-frequency expressions in terms of spectral weight. In this section, we obtain results for the electrical resistivity $\rho$ within IPT-$D$ to verify whether they are physically sensible at strong coupling. With IPT-$n_0$ they are not: One obtains an insulator at finite filling on the electron-doped side. Since vertex corrections vanish in single-site DMFT, the conductivity can be obtained from
\begin{equation}\label{conduc_1}
    \sigma_{xx}(0) = \sigma_0\int d\omega\left(-\d{\pd f(\omega)}{\pd \omega}\right)\sum_k\left(\d{\pd\varepsilon_k}{\pd k_x}\right)^2A^2(k,\omega),
\end{equation}
where $\sigma_0 = \d{\pi e^2}{\hbar a}$ and $A(k,\omega) = -\d{1}{\pi}\text{Im}\left\{\d{1}{\omega - (\varepsilon_k-\mu) - \Sigma (\omega)}\right\}$. We have restored the lattice spacing $a$ to exhibit the units. Because the self-energy is local, we can also replace the triple sum over $k$ by a one dimensional integral over an energy variable.
\begin{equation}\label{conduc_2}
    \sigma_{xx}(0) = \sigma_0 \int d\varepsilon X(\varepsilon)\int d\omega\left(-\d{\pd f(\omega)}{\pd \omega}\right)A^2(\varepsilon,\omega),
\end{equation}
where the so called transport function that includes the effect of the lattice is $X(\varepsilon) = \sum_k\left(\d{\pd\varepsilon_k}{\pd k_x}\right)^2\delta(\varepsilon-\varepsilon_k)$. For a simple cubic lattice in any dimension, the calculation of this function can be brought in the form of a one-dimensional integral\cite{Arsenault:2011}, but on the FCC  lattice this is not possible. If one wishes to use the expression Eq.~\eqref{conduc_2} instead of keeping the full three dimensional integrals in Eq.~\eqref{conduc_1}, $X$ must be calculated numerically. We explain in Appendix~\ref{MC_function} an efficient way to do this. The result for the FCC lattice is shown in Fig.~\ref{fig:X_func}. We have tested both ways of obtaining $\sigma_{xx}(0)$, i.e. Eq.~\eqref{conduc_2} and Eq.~\eqref{conduc_1} and they give essentially the same answer. They cannot give exactly the same number because $X(\varepsilon)$ is calculated for a fixed number of points and we must thus interpolate between these points when performing the integral over $\varepsilon$. This adds another source of numerical error not present in Eq.~\eqref{conduc_1}. But, contrary to the non-interacting density of states (Fig.~\ref{fig:DOS_U0}), $X(\varepsilon)$ is a smooth function (Fig.~\ref{fig:X_func}) and thus the process of interpolation will only induce a negligible error. When the integrals are performed in the order shown in Eq.~\eqref{conduc_2}, i.e. integrate over $\omega$ first, the resulting integrand for the $\varepsilon$ integration is smooth and $X(\varepsilon)$ need not be obtained with extreme accuracy. For these reasons, for the conductivity we preferred to use Eq.~\eqref{conduc_2}. Using Eq.~\eqref{conduc_1} increases dramatically the calculation time here contrary to $G(i\omega_n)$ or $\chi_{11}(i\Omega_n)$ discussed in the next section.

At low temperature and finite doping, the conductivity is proportional to $X(\tilde{\mu})$. Thus, if $\tilde{\mu}$ is such that its value is outside the non-interacting band $-12\ldots 4$, $X(\tilde{\mu}) = 0$ and thus $\sigma_{xx}(0) = 0$. In an exact implementation of single-site DMFT, a null conductivity can only happen at half-filling for $U > U_{Mott}$. At any finite doping, there is a quasi-particle peak and Luttinger's theorem is respected (Fig.\ref{fig:Fermi_CTQMC}-(a)). As IPT-$n_0$ fails with respect to Luttinger's theorem (Fig.\ref{fig:Fermi_IPT}-(a)) in that case $\rho$ start diverging at low $T$ for the densities with $\tilde{\mu}$ outside the band while it should exhibit the $T^2$ behavior of a Fermi liquid.

We show in Fig.\ref{fig:resis} the result for IPT-$D$. We see that it has the correct $T^2$ behavior at low temperature. For clarity, we omitted values of densities between $n = 0.80$ and $n = 1.2$. Nothing very different happens there. We still have the low temperature $T^2$ behavior and, as is already obvious from the figure, the absolute values of $\rho$ obtained are larger and larger when we approach half-filling, the Mott insulating state. Note that close to half-filling, even though there is a tendency for the resistivity to saturate at very high temperature \cite{Jarrell:1994}, this occurs at values of the resistivity much larger than the Mott-Ioffe-Regel limit $\rho \sim \hbar a/e^2$. This is characteristic of incoherent transport in strongly correlated systems.

\subsection{Optical conductivity}

\begin{figure}[tbp]
\begin{center}
\includegraphics[width=0.95\linewidth]{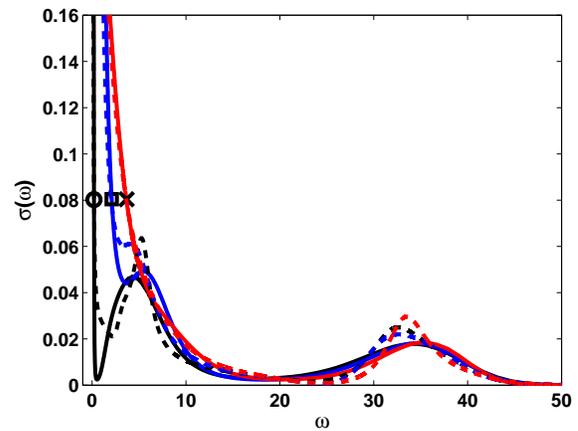}
\end{center}
\caption{(Color online) Optical conductivity for $U = 32t$ and $n = 0.80$, as calculated from IPT-$D$ for three different temperatures using two analytical continuation approaches for each temperature. The maximum entropy results are represented with solid lines and the Pad\'{e} analytical continuations with dashed lines. The broadest zero-frequency peak (red ($\times$)) is for the largest temperature, $\beta = 1/t$, and the narrowest one (black ($\circ$)) for the lowest temperature, $\beta = 25/t$. The intermediate case, $\beta = 2.3/t$ is in blue ($\square$). The features in the optical conductivity can be identified with transitions between the Fermi level and peaks in the single-particle density of states.}
\label{fig:opt_conduc}
\end{figure}

We analyzed the performance of our new solver IPT-$D$ for Fermi surface properties at $T = 0$ ($Z$,$\Sigma'$,$\Sigma''$), for integrated quantities like chemical potential $\mu$, energy, specific heat, resistivity and frequency-dependent functions such as the density of states. To finish, we look at the optical conductivity. Appendix.~\ref{appen_opt_conduc} explains how it is calculated using the susceptibility $\chi_{11}(i\Omega_n)$ in bosonic Matsubara frequency and analytical continuation.

As a first check, which does not depend on analytical continuation, we verify the $f$-sum rule $\chi_{11}(i\Omega_n=0) = \sum_k\d{\pd^2\varepsilon_k}{\pd k_x^2}\langle n_k\rangle$ (see for example Ref.~[\onlinecite{Bergeron:2011}]). In the case of nearest-neighbor hopping, this quantity can be related to the kinetic energy. For a FCC lattice one finds $\sum_k\d{\pd^2\varepsilon_k}{\pd k_x^2}\langle n_k\rangle = -\d{2}{3}\langle K \rangle$ where $\langle K \rangle$ is the average kinetic energy. For CTQMC the average kinetic energy is obtained from $\langle K \rangle = -T\langle k \rangle$ where, as already mentioned, $\langle k \rangle$ is the average perturbation order obtained directly from the Monte-Carlo simulation. To calculate $\chi_{11}(i\Omega_n=0)$ one needs the dressed Green's function, or equivalently the self-energy ($\Sigma (i\omega_n)$). A very large number of numerical operations is necessary, as explained in Appendix.~\ref{appen_opt_conduc}, but analytical continuation is unnecessary. For all the tests we did, the ratio $\left|\d{\chi_{11}(i\Omega_n=0)}{\langle K \rangle}\right|$ agreed with $2/3$ up to the third digit. For example, for $n = 0.84$, $\beta/t = 25$ and $U=32t$ the ratio is 0.6668, while for $n=0.80$ it is 0.6664.

Analytical continuation of CTQMC is problematic, especially since we have a very wide frequency range given the large value of $U$. Hence we display only results obtained with analytical continuation of IPT-$D$ and check for consistency with what is expected from the density of states. Results for density $n = 0.80$, interaction $U=32$ and three temperatures is shown in Fig.~\ref{fig:opt_conduc}. Solid and dashed lines correspond respectively to Maximum Entropy \cite{Bergeron:2011} and Pad\'e analytical continuation. There is some quantitative disagreement but the qualitative information is the same. At low temperature ($\beta/t = 25$ black lines) there is a clear peak around $\omega \approx 5$ which corresponds to transitions between the lower Hubbard band and the quasi-particle peak appearing in the density of state for a similar density in Fig.\ref{fig:DOS}. At larger temperature (blue lines $\beta/t = 2.3$), the decrease of the peak near $\omega \approx 5$ in $\sigma (\omega)$ corresponds to the disappearance of the quasi-particle peak and loss of coherence. That loss of coherence for $n=0.80$ is signaled by the maximum in $C_n$ observed in Fig.~\ref{fig:Cv}. At an even larger temperature, (red lines $\beta/t = 1$) the $\omega \approx 5$ peak has disappeared since we are now in the incoherent regime. The peak for transition to the upper Hubbard band around $32t$ is always visible.

\section{Conclusion}

In addition to being numerically inexpensive, IPT provides a method where analytically continued results can be reliably obtained directly in real frequencies or from Pad\'e approximants\cite{Vidberg:1977} instead of Maximum Entropy methods required when Quantum Monte Carlo is used as an impurity solver.

However, for large interaction strengths in doped Mott insulators, the popular condition for IPT where one imposes $n=n_0$ fails at low temperature for a broader regime than previously expected\cite{Potthoff:1997}. As a solution, we propose that one should instead enforce the exact relation Eq.\eqref{D_occ_sigG} between double occupancy and single particle quantities. Further improvements are expected if one also enforces the third moment of the spectral weight\cite{Potthoff:1997}.

Our new method, IPT-$D$, can be used for any coupling if $D$ is known. Double occupancy $D$ can be obtained quite accurately by a number of methods and is negligibly dependent on temperature for large coupling. For example, in the strong coupling regime one can use exact diagonalization of small clusters or slave bosons\cite{Georges:1996} while, at weak coupling, methods such as Two-Particle-Self-Consistent theory (TPSC)\cite{Vilk:1997,TremblayMancini:2011} give good results. Also, in both regimes, Quantum Monte-Carlo methods can be used. In the very large $U$ limit, the naive estimate $D=0$ for $n<1$ and $D=n-1$ for $n>1$ leads to qualitatively correct results, except very close to half-filling. Our approach has been benchmarked on the FCC lattice for a number of observables. The large-particle hole asymmetry of that lattice survives in observable quantities even for interaction strength equal to twice the bandwidth.

\begin{acknowledgments}
The authors thank Dominic Bergeron for the maximum entropy
codes that we used for analytical continuation of the CTQMC data and for discussions. This work was partially supported by NSERC (L.-F.A. and A.-M.S.T.), the Tier I Canada Research Chair Program (A.-M. S. T.), and Universit\'{e} de Sherbrooke. A.-M.S.T is grateful to the Harvard Physics Department for support and P. S\'emon for hospitality during the writing of this work. Partial support was also provided by the MIT-Harvard Center for Ultracold Atoms. Simulations were performed using a code based on the ALPS library~\cite{ALPS} on computers provided by CFI, MELS, Calcul Qu\'ebec and Compute Canada. Portions of the hybridization expansion impurity solver developed by P. S\'{e}mon were inspired by the code gracefully provided by E. Gull and P. Werner.
\end{acknowledgments}

\appendix
\section{Integrator}\label{appen_integrate}
\begin{figure}[tbp]
\begin{center}
\mbox{\includegraphics[width=0.9\linewidth]{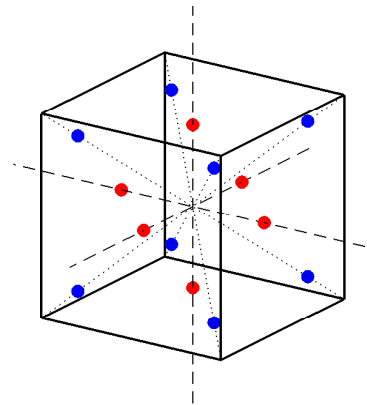}}
\end{center}
\caption{The special points for a 3d Gaussian quadrature of fifth order over a cube of length $2h$.}
\label{fig:cube}
\end{figure}
As explained  in the main text, for accuracy in the DMFT iteration we need to perform integrals over a three dimensional Brillouin zone. In this appendix we use symmetry and ideas from Gaussian quadrature, adaptive methods, and statistics, to devise an accurate and fast integrator. We first obtain a quadrature of order five and then explain how we can make it adaptive.

We need a normalized triple integral over a cube of length $2h$ centered at a point $\textbf{r}_0 = [x_0,y_0,z_0]$. If we put the origin at this point, the integral takes the form
\begin{equation}\label{triple_int}
    \d{1}{(2h)^3}\int_{-h}^h\int_{-h}^h\int_{-h}^hdxdydz f(x,y,z).
\end{equation}
We considered a normalized integral because the integrals we need to solve are over k-space and thus need normalization.

We first show that fourteen appropriately chosen points and only two weights can give us an approximation of order five. Usually when one develops a Gaussian quadrature, only the order is specified and the points and weights are obtained. In 3d, this may be very cumbersome so we start with points symmetrically placed and we will show that they give a good approximation. Take points on each axis and on the diagonals of the cube, $(\pm a,0,0)$, $(0,\pm a,0)$ , $(0,0,\pm a)$ and $(\pm b, \pm b, \pm b)$ as illustrated in Fig.~\ref{fig:cube}. By symmetry, there are only two weights $w_1$ and $w_2$. Thus the integral is approximated by
\begin{equation}\label{triple_int_gauss_1}
\begin{split}
    &\d{1}{(2h)^3}\int_{-h}^h\int_{-h}^h\int_{-h}^hdxdydz f(x,y,z)\\ &\approx w_1\left[ f(\pm a,0,0) + f(0,\pm a,0) + f(0,0,\pm a)\right]\\ &+ w_2f(\pm b, \pm b, \pm b).
\end{split}
\end{equation}
To determine the numbers $a$, $b$, $w_1$ and $w_2$ we require that every polynomial of order five or less should be integrated exactly by this scheme. In 3d, this corresponds to many different polynomials but we have only four unknowns and apparently too many equations. This is where symmetry comes into play. First, since we integrate over a cube from $-h$ to $h$, all odd polynomials integrate to zero. Also, for example, a polynomial of the form $x^2y^2$ is equivalent to $y^2z^2$, or $x^2$ is equivalent to $y^2$ and $z^2$ and so on for every type of polynomials. Thus we only have to consider four different polynomials i.e. $1$, $x^2$, $x^4$ and $x^2y^2$.\\
\\
Taking $f(x,y,z) = 1$, the integral gives one and thus we obtain the first equation
\begin{equation}\label{gauss_eq1}
    6w_1 + 8w_2 = 1
\end{equation}
Taking $f(x,y,z) = x^2$, the integral gives $\d{1}{(2h)^3}\int dxdydz x^2 = \d{h^2}{3}$ and we obtain
\begin{equation}\label{gauss_eq2}
    2a^2w_1 + 8b^2w_2 = \d{h^2}{3}
\end{equation}
Similarly, with $f(x,y,z) = x^4$, we have $\d{1}{(2h)^3}\int dxdydz x^4 = \d{h^4}{5}$ and we find
\begin{equation}\label{gauss_eq3}
    2a^4w_1 + 8b^4w_2 = \d{h^4}{5}
\end{equation}
Finally, taking $f(x,y,z) = x^2y^2$, $\d{1}{(2h)^3}\int dxdydz x^2y^2 = \d{h^4}{9}$ and we obtain
\begin{equation}\label{gauss_eq4}
    8b^4w_2 = \d{h^4}{9}
\end{equation}
Solving these equations, we obtain
\begin{equation}\label{param_Gauss}
\begin{split}
    w_1 &= \d{40}{361}\\
    w_2 &= \d{121}{2888}\\
    a &= \sqrt{\d{19}{30}}h\\
    b &= \sqrt{\d{19}{33}}h
\end{split}
\end{equation}

To make the method adaptive, we take our cube and split it in eight. If we take one of these cubes, and put the origin in its center we now have the integral over a cube from $-\d{h}{2}$ to $\d{h}{2}$ centered at $\textbf{r}_0$. We can thus use Eq.~\eqref{triple_int_gauss_1} but with $h \rightarrow \d{h}{2}$ and the points $(x_0\pm a,y_0,z_0)$, $(x_0,y_0\pm a,z_0)$, $(x_0,y_0,z_0\pm a)$ and $(x_0\pm b,y_0\pm b,z_0\pm b)$. We can do this for each of the eight cubes obtaining the new approximation for the integral $I = \d{1}{8}\sum_iI_i$. The process can be repeated. Each of the eight cubes can be subdivided again with integrals from $-\d{h}{4}$ to $\d{h}{4}$. When one subdivision has converged, this part is stopped. The calculation has converged when all subdivisions have converged.

The convergence criterion requires a detailed discussion. Assume that we aim at a relative error $\epsilon$. For an adaptive method, we need absolute error. Indeed, with a simple 1d adaptive integration method where one subdivides the interval in half, if one wishes an absolute error $\delta$ one usually imposes absolute error $\d{\delta}{2}$ on each of the two sub-intervals. To determine the absolute error in our case, we first estimate the value of the integral by performing three subdivisions, i.e. using $8^3=512$ cubes. Let us call the resulting integral $I^{(3)}$. Then we take for the absolute error needed for the adaptive integration method $\delta = \epsilon I^{(3)}$. If we imagine launching the integrator from scratch, one would ask for $\delta/8$ accuracy in each of the $8$ sub-cubes when we do a division. This often leads to a final answer that is more accurate than desired. For heavy numerical calculations it is desirable to optimize the choice of the error in each subinterval to minimize the computation time while maintaining the final desired accuracy. In our case, we claim that it suffices to require the absolute error within each subinterval to be $\d{\delta}{\sqrt{8}}$ instead of $\d{\delta}{8}$, as we might have naively expected. Indeed, if we consider each value on the sub-cube as a random variable $I_i$ and want an error $\delta$ on the original cube i.e. on the sum $I=\sum_iI_i$, an error $\d{\delta}{\sqrt{8}}$ for each sub-cube suffices is we assume that the errors on the $I_i$'s are independent and uniformly distributed (IUD). Indeed, in that case $\text{Var}(I) = \sum_i\text{Var}(I_i)$ and thus $\delta_I = \sqrt{8}\delta_{I_i}$. This hypothesis of an IUD is of course not rigorous, but we have extensively tested this choice for the error with many different integrands with known integrals. By taking advantage of the fact that statistical hypothesis on the errors become reasonable since the integral is high dimensional, our approach is faster. We also checked that our approach is more precise and faster than using three adaptive 1d integrators. Additional speedup can be obtained by taking into account the symmetry of the integrand.
\section{IPT-$D$ implementation}\label{appen_IPT}
\begin{figure}[tbp]
\begin{center}
\includegraphics[scale=0.27,angle=0]{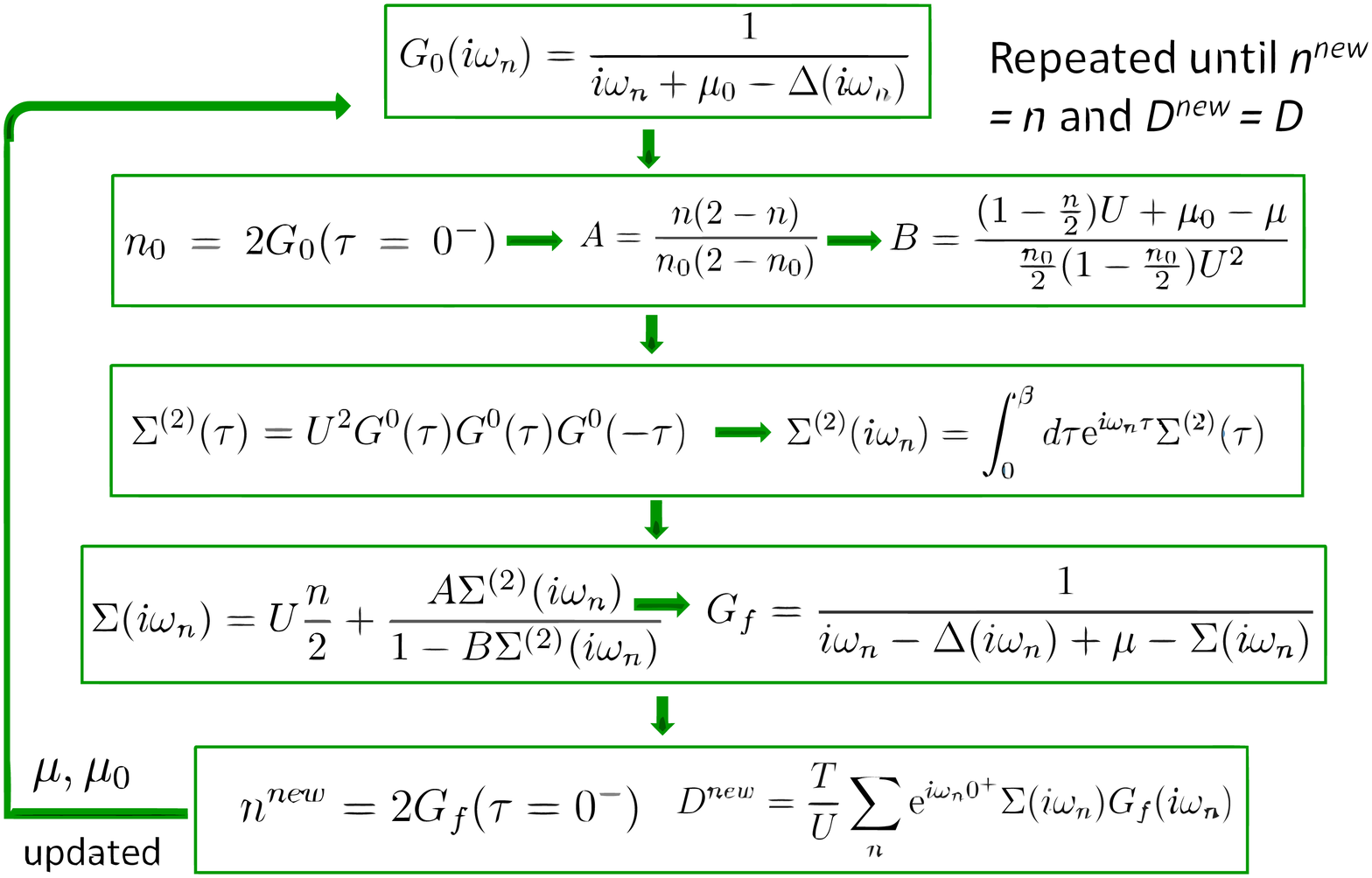}
\end{center}
\caption{Flow chart for the impurity solver loop in IPT. There is also an outer loop for $\Delta$, see text. }
\label{fig:loop_IPT}
\end{figure}
In this section we detail how we implemented IPT-$D$. IPT as an approximative solver is fast, but it needs to also be implemented in the fastest possible way. In IPT, we are solving a system of two nonlinear equations with two unknowns: $\mu_0$ and $\mu$. The equations are $n-2\d{1}{\beta}\sum_n\text{e}^{i\omega_n0^+}G(i\omega_n)=0$ and $D - \d{T}{U}\sum_n\text{e}^{i\omega_n0^+}\Sigma (i\omega_n)G(i\omega_n) = 0$ where $n$ and $D$ are fixed numbers for a particular set of parameters. The self-energy must be calculated using Eqs.~\eqref{IPT_self},\eqref{self_2order_AIM}. We show how to do this efficiently.

We first start with guesses for the hybridization function $\Delta (i\omega_n)$, and the two chemical potentials $\mu_0$ and $\mu$. Then we calculate the impurity model loop as shown in Fig.~\ref{fig:loop_IPT}. Once the loop has been converged, we use the self-energy to calculate the lattice Green's function $G(i\omega_n) = \sum_k\d{1}{i\omega_n-(\varepsilon_k-\mu)-\Sigma (i\omega_n)}$. With it, the new hybridization function can be calculated $\Delta (i\omega_n) = -\Sigma (i\omega_n) - G^{-1}(i\omega_n) + i\omega_n + \mu$ and finally $\Delta$, $\mu$ and $\mu_0$ are fed back to the impurity model loop. This is repeated until global convergence is reached.

It must be specified here that $\Delta$ is a function of both $\mu$ and $\mu_0$ but in the present algorithm, once $\Delta$ is fed to the impurity loop, it is considered to be independent while $\mu$ and $\mu_0$ are iterated until we obtain the correct $n$ and $D$. However, once the system is close to convergence, the difference between $\Delta$ in the inner loop and the correct $\Delta (\mu,\mu_0)$ becomes really small and once convergence is reached, it is indeed the same function. This is also the approach that was adopted originally in [\onlinecite{Kajueter:1996}]. This approach is much faster than fixing $\mu$ and $\mu_0$, converging the entire DMFT calculation, calculating the new $n$ and $D$, updating $\mu$ and $\mu_0$ and converging again the DMFT calculation until we obtain the correct $n$ and $D$. We have tested and used the two methods and they give the same results. The fast method can, for some particular parameter set, become unstable and thus, in these cases, the long, more rigorous method, may be used.

To perform this calculation, we see from Fig.~\ref{fig:loop_IPT} that we must calculate Fourier transforms from Matsubara frequencies to imaginary time and back. This must be calculated numerically and we now explain how to do it efficiently. The first necessary step is to calculate $G_0(\tau)$. We can show that the function $\Delta (i\omega_n)$ for a one band model with a dispersion relation $\varepsilon_k$ behaves asymptotically like $\Delta (i\omega_n)_{n\rightarrow \infty} \rightarrow \d{1}{i\omega_n}\sum_k\varepsilon_k^2 \equiv \d{c}{i\omega_n}$, similar to what was previously obtained\cite{Koch:2008}. For the 3d FCC lattice with nearest-neighbor hopping $c = 12t^2$. We can use this to define $G_0^{inf}(i\omega_n) = \d{1}{i\omega_n + \tilde{\mu}_0 - \d{c}{i\omega_n}}$ that gives the asymptotic high-frequency behavior of $G_0$. We need it because the Fourier transform necessary to get $G_0(\tau)$ must be approximated by a finite sum and thus the function must be convergent at least as fast as $\d{1}{(i\omega_n)^2}$, while $G_0(i\omega_n) \rightarrow \d{1}{i\omega_n}$. We thus consider the function $F = G_0 - G_0^{inf}$ instead and add the missing terms analytically.\\
\\
\begin{equation}\label{G_0_tau_1}
\begin{split}
    G_0(\tau) &= T\sum_{-N/2}^{N/2 -1}\text{e}^{-i\pi (2n+1)j/N}F(i\omega_n)\\
    &+ T\sum_n\text{e}^{-i\omega_n\tau}G_0^{inf}(i\omega_n)\\
    &= \text{e}^{-i\pi j(1/N -1)}T\underbrace{\sum_{n=0}^{N-1}\text{e}^{-i2\pi nj/N}F(i\omega_{n-N/2})}_{\text{definition of FFT}}\\ &+ T\sum_n\text{e}^{-i\omega_n\tau}G_0^{inf}(i\omega_n).
\end{split}
\end{equation}
In the first sum, imaginary time has been discretized in $N$ bins so that $j/N=\tau/\beta$. There is then a maximum and minimum Matsubara frequency. In the second equality of Eq.~\eqref{G_0_tau_1}, FFT stands for Fast Fourier Transform. The last term, $T\sum_n\text{e}^{-i\omega_n\tau}G_0^{inf}(i\omega_n)$ can be calculated analytically using complex analysis. We find, for $\tau>0$,
\begin{equation}
\begin{split}
    T\sum_n\text{e}^{-i\omega_n\tau}G_0^{inf}(i\omega_n) = &-\d{z_1}{z_1-z_2}f(-z_1)\text{e}^{-z_1\tau}\\ &-\d{z_2}{z_2-z_1}f(-z_2)\text{e}^{-z_2\tau} \equiv H^>(\tau),
\end{split}
\end{equation}
where $f(z)$ is the Fermi function and $z_j = \d{-\mu_0 \pm \sqrt{ \mu_0^2 + 4c }}{2}$.
Hence, in terms of FFT's, we obtain
\begin{equation}\label{G_FFT_ptau}
    G_0(\tau) = \text{e}^{-i\pi j(1/N -1)}\d{1}{\beta}\text{FFT}(F(i\omega_{n-N/2})) + H^>(\tau).
\end{equation}
We must remember that FFT does not give the value at $\tau = \beta$. We will come back to that point later. We also need $G_0(-\tau)$. This can be done using the antiperiodic property or by using a procedure similar to $G_0(\tau)$. In that case, we obtain (again for $\tau>0$)

\begin{equation}\label{G_FFT_mtau}
    G_0(-\tau) = \text{e}^{-i\pi j(1/N -1)}\d{1}{\beta}\text{FFT}(F^*(i\omega_{n-N/2})) + H^<(\tau),
\end{equation}
where
\begin{equation}
H^<(\tau) \equiv \d{z_1}{z_1-z_2}f(z_1)\text{e}^{z_1\tau} + \d{z_2}{z_2-z_1}f(z_2)\text{e}^{z_2\tau}.
\end{equation}

With these two Green's function we can calculate the second order contribution to the AIM self-energy that appears in IPT Eq.~\eqref{self_2order_AIM}. In this equation, we need to perform an integral. To obtain an asymptotic behavior in Matsubara frequencies that decays instead of being periodic, we cannot do a direct integration\cite{Georges:1996}. The trick here is to perform a cubic spline interpolation of $I(\tau)= G_0^2(\tau)G_0(-\tau)$ and then Fourier transform that spline interpolation. To obtain the spline, one needs two conditions to solve the system of equations. In our case, it suffices to find the derivatives at $\tau = 0^+$ and $\tau = \beta^-$. We will show later that making the Fourier transform of the spline is really accurate and introduces a minimum of numerical errors.

Up to now, we have only considered $\tau > 0$ and thus what we need are the derivative at $0^+$ and at $\beta^-$. If, for the moment, we consider a paramagnetic system we can write $I(\tau) = G_0^2(\tau)G_0(-\tau)$. The derivative is thus
\begin{equation}\label{dF_dtau}
    \d{dI(\tau)}{d\tau} = 2G_0(\tau)G_0(-\tau)\d{dG_0(\tau)}{d\tau} + G_0^2(\tau)\d{dG_0(-\tau)}{d\tau}.
\end{equation}
We calculate the derivative of the Green's functions from their definition
\begin{equation}
\begin{split}
    \d{dG_0(\tau)}{d\tau} &= T\sum_n\text{e}^{-i\omega_n\tau}(-i\omega_n)F(i\omega_n) + \d{dH^>(\tau)}{d\tau}\\
    \d{dG_0(-\tau)}{d\tau} &= T\sum_n\text{e}^{-i\omega_n\tau}(-i\omega_n)F^*(i\omega_n) + \d{dH^<(\tau)}{d\tau}.
\end{split}
\end{equation}
By defining $F_1(i\omega_n) \equiv (-i\omega_n)F(i\omega_n)$ and $F_2(i\omega_n) \equiv (-i\omega_n)F^*(i\omega_n)$, we obtain
\begin{equation}
\begin{split}
    \d{dG_0(\tau)}{d\tau} &= \text{e}^{-i\pi j[1/N - 1]}\d{1}{\beta}\text{FFT}(F_1(i\omega_n))\\ &+\d{z_1^2}{z_1-z_2}f(-z_1)\text{e}^{-z_1\tau} +\d{z_2^2}{z_2-z_1}f(-z_2)\text{e}^{-z_2\tau} \\
    \d{dG_0(-\tau)}{d\tau} &= \text{e}^{-i\pi j[1/N - 1]}\d{1}{\beta}\text{FFT}(F_2(i\omega_n))\\ &+ \d{z_1^2}{z_1-z_2}f(z_1)\text{e}^{z_1\tau} + \d{z_2^2}{z_2-z_1}f(z_2)\text{e}^{z_2\tau}.
\end{split}
\end{equation}
With these two equations, we can get the derivatives at $\tau = 0^+$. To obtain the Green's function and its derivative at $\tau = \beta^-$, we use the spectral representation of the Green's function to show that $G_0(\beta^-) = -1 - G_0(0^+)$, $\d{dG_0(\tau)}{d\tau}\Big|_{\tau = \beta^-} = -\tilde{\mu}_0 - \d{dG_0(\tau)}{d\tau}\Big|_{\tau = 0^+}$ and $\d{dG_0(-\tau)}{d\tau}\Big|_{\tau = \beta^-} = -\tilde{\mu}_0 - \d{dG_0(-\tau)}{d\tau}\Big|_{\tau = 0^+}$.

We now have everything we need to calculate the derivative of Eq.~\eqref{dF_dtau} for $\tau = 0^+$ and $\tau = \beta^-$. Knowing $I(\tau)$ and $\d{dI(\tau)}{d\tau}$ we can calculate the coefficients of the spline. We need the Matsubara-Fourier transform of functions represented by that spline on the right-hand side of
\begin{equation}\label{FT_t_wn}
    f(i\omega_m) = \int_{\tau_0}^{\tau_N}d\tau\text{e}^{i\omega_m\tau}f(\tau).
\end{equation}
Let us call $S(\tau)$ the piecewise cubic spline for $f(\tau)$. The method is presented in details in Appendix E of [\onlinecite{Bergeron:2011}]. Integrating by parts, the result for fermionic frequencies is
\begin{equation}\label{f_fermion}
\begin{split}
    f(i\omega_m) = &\d{-S_1(0) - S_N(\beta)}{i\omega_m} + \d{S_1'(0) + S_N'(\beta)}{(i\omega_m)^2}\\ &+ \d{-S_1''(0) -S_N''(\beta)}{(i\omega_m)^3}\\ &+ N\d{\left(1-\text{e}^{i\omega_m\d{\beta}{N}}\right)}{(i\omega_m)^4}\text{IFFT}\left(\text{e}^{\d{i\pi n}{N}}S_{n+1}'''\right),
\end{split}
\end{equation}
where IFFT is the inverse Fast Fourier transform. This result is what we needed since it has the correct high-frequency behavior where in principle the first three terms are exact while the last one is obtained from a numerical inverse Fourier transform. Since the latter is the coefficient of $\d{1}{(i\omega_n)^4}$, errors do not adversely affect the high-frequency behavior.

\section{Calculation of $N_0(\varepsilon)$ and $X(\varepsilon)$}\label{MC_function}
Even though for the DMFT iterations we found that the way to obtain accurate results was to use the adaptive method described in Appendix \ref{appen_integrate}, we show here how to calculate $N_0(\varepsilon)$, the non-interacting density of states and $X(\varepsilon)$. The latter quantity appears in calculations of the conductivity and transport properties in general and it is in this context that we used the results presented here. Both quantities have the general form $\sum_kF(\textbf{k})\delta (\varepsilon-\varepsilon_f)$. Since our band structure is not simple, we cannot perform the integral analytically. The question is thus, how do we treat the delta function in a numerical calculation?\\
\\
A simple approach would be to replace the delta function by a Lorentzian and perform the integral using an adaptive scheme. But, this approximation for the delta function gives tails at the edges of the band. A Monte Carlo scheme is preferable not only because the sharpness of the delta function is maintained without tails, but also because one does not need to do a triple integral for each value of $\varepsilon$: The complete function of $\varepsilon$ is be obtained at once.\\
\\
\begin{figure}[tbp]
    \begin{center}
        \includegraphics[scale=0.5,bb= 132 370 512 571]{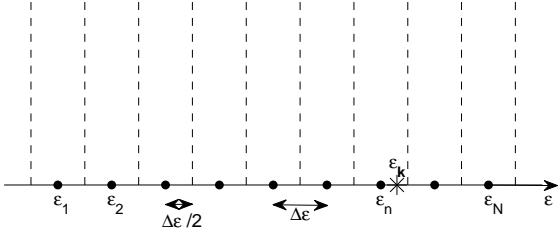}
    \end{center}
    \caption{Illustration of the Monte Carlo integration scheme used to obtain $N_0(\varepsilon)$ and $X(\varepsilon)$}
  \label{fig:MC_scheme}
\end{figure}
We first choose for how many energy $\varepsilon$ points we want to know the function. This number defines a number of bins, shown as dashed lines in Fig.~\ref{fig:MC_scheme}. One then generates a random point $(k_x,k_y,k_z)$, calculates $\varepsilon_k$ and locates the bin where this number belongs. For example, in Fig.~\ref{fig:MC_scheme}, the random $\varepsilon_k$ belongs to the bin $n$. We then add to this bin $\d{1}{\Delta\varepsilon}F(\textbf{k})$. This is equivalent to approximating the delta function by a rectangle of finite width. We continue this process $M$ times and divide, at the end, the numbers in the bins by $M$. The function is thus given by the numbers in the bins, each bin corresponding to a particular energy $\varepsilon$. Accuracy and smoothness can be improved by increasing the number of random points.
\section{Optical conductivity}\label{appen_opt_conduc}
For the optical conductivity, we need the current-current correlation function in bosonic Matsubara frequencies. In DMFT, vertex corrections vanish\cite{Georges:1996}, hence we have
\begin{equation}\label{chi_11_def}
    \chi_{11}(i\Omega_l) = -\sum_{k,\sigma}v_k^2\d{1}{\beta}\sum_{\omega_n}G_{\sigma}(k,i\omega_n)G_{\sigma}(k,i\omega_n+i\Omega_l),
\end{equation}
where $\Omega_l$ are bosonic Matsubara frequencies while $\omega_n$ are fermionic.
To compute this, we again use the convolution theorem and FFT, as described for IPT in Appendix \ref{appen_IPT}. The above equation can be written as
\begin{equation}\label{chi_11_2}
\begin{split}
    \chi_{11}(i\Omega_l) &= -\int_0^{\beta} d\tau\text{e}^{i\Omega_l\tau}\sum_{k,\sigma}v_k^2G_{\sigma}(k,\tau)G_{\sigma}(k,-\tau)\\
    &= -\int_0^{\beta} d\tau\text{e}^{i\Omega_l\tau}F(\tau).
\end{split}
\end{equation}
 The only difference with Eq.~\eqref{FT_t_wn} is that here we have bosonic frequencies. Using again the cubic spline trick we obtain an expression similar to Eq.~\eqref{f_fermion}. For $\Omega_l \neq 0$
\begin{equation}
\begin{split}
    f(i\Omega_l) &= \d{-S_1(0) + S_N(\beta)}{i\Omega_l} + \d{S_1'(0) - S_N'(\beta)}{(i\Omega_l)^2}\\ &+ \d{-S_1''(0) + S_N''(\beta)}{(i\Omega_l)^3}+ N\d{\left(1-\text{e}^{i\Omega_l\d{\beta}{N}}\right)}{(i\Omega_l)^4}\text{IFFT}\left(S_{n+1}'''\right).
\end{split}
\end{equation}
while for $\Omega_l = 0$
\begin{equation}
\begin{split}
    f(0) = \sum_{n=1}^N\Big[& \d{a_n}{4}\left(\tau_{n}^4-\tau_{n-1}^4\right) + \d{b_n}{3}\left(\tau_{n}^3-\tau_{n-1}^3\right)\\ &+ \d{c_n}{2}\left(\tau_{n}^2-\tau_{n-1}^2\right) + d_n\left(\tau_{n}-\tau_{n-1}\right) \Big].
\end{split}
\end{equation}
with $a_n$, $b_n$, $c_n$ and $d_n$ the coefficients of the cubic spline.

We also need to go from Matsubara frequencies to imaginary time for $G_k(\tau)$ and for $F(\tau)$ in  Eq.~\eqref{chi_11_2}. We proceed to obtain the analog of Eq.~\eqref{G_FFT_ptau}. $G_{inf}$ has a similar structure since the asymptotic behavior of the self-energy is $\Sigma (i\omega_n) = Un/2 + \d{U^2n(2-n)}{4i\omega_n}$. This time the poles are at
\begin{equation}\label{poles}
    z_j = \d{(\varepsilon_k-\mu+Un/2) \pm \sqrt{ (\varepsilon_k-\mu+Un/2)^2 + 4c }}{2},
\end{equation}
where $c = U^2n(2-n)/4$.\\
\\
Once again, the values at $\tau=\beta^-$ are found using the spectral representation of $G_k$. The expressions are
\begin{equation}\label{G0beta1}
    G_k(\beta^-) = -1-G_k(0^+).
\end{equation}
\begin{equation}\label{Gm0beta1}
    G_k(-\tau)\Big|_{\tau = \beta^-} = 1-G_k(-\tau)\Big|_{\tau = 0^+}.
\end{equation}
\begin{equation}\label{dG0beta2}
    \d{dG_k(\tau)}{d\tau}\Big|_{\tau=\beta^-} = (\varepsilon_k - \mu +Un/2) - \d{dG_k(\tau)}{d\tau}\Big|_{\tau=0^+}.
\end{equation}
\begin{equation}\label{dGm0beta2}
    \d{dG_k(-\tau)}{d\tau}\Big|_{\tau=\beta^-} = (\varepsilon_k - \mu +Un/2) - \d{dG_k(-\tau)}{d\tau}\Big|_{\tau=0^+}.
\end{equation}
We can thus calculate $F(\tau)$ appearing in Eq.~\eqref{chi_11_2} and its first derivative at $\tau = 0^+$ and $\tau = \beta^-$ to find the cubic spline interpolation.

The sum over wave vectors $k$ requires some comments. Like for the DMFT calculation, we use the adaptive scheme in Appendix \ref{appen_integrate}. However, if we look at $F(\tau)$ in Eq.~\eqref{chi_11_2}, in principle for each $\tau$ we need to perform the integral over $k$ independently, but the whole point of FFT is to obtain all $\tau$ points at the same time. Our solution is to launch the integrator for all $\tau$ at the same time and keep in memory the estimate of the integral for each $\tau$ while we refine the estimate. Once the integrator converges for $\tau = 0$, we conclude that this $k$ grid is the one for all $\tau$ and stop the calculation. Since we keep all values in memory, we have the function for all $\tau$ calculated for the grid in $k$ space appropriate for $\tau = 0$. We verified that by converging the calculation for other values of $\tau$ we obtain the same answer. We have also tried the other way around, where we interchange the integral over $k$ and the Fourier transform in Eq.~\eqref{chi_11_2}. In this case, we converge the zero frequency and, once again, the results are essentially the same.

Once we have calculated $\chi_{11}(i\Omega_l)$, we need to obtain the real frequency representation since the optical conductivity is given by $\d{\chi_{11}''(\omega)}{\omega}$. In the case of CTQMC, we use the maxent analytical continuation scheme developed in Bergeron \emph{et al.} [\onlinecite{Bergeron:2011}] for the conductivity. For IPT results we use both Pad? and maxent to find $\chi_{11}(\omega)$.




\end{document}